\documentclass[useAMS,usenatbib]{mn2e}
\usepackage{graphicx}
\usepackage{color}
\usepackage{soul}
\usepackage{bm}
\usepackage{natbib}
\usepackage{amsmath}	
\usepackage{amssymb}	

\newcommand{\bec}[1]{\mbox{\boldmath $ #1$}}

\newcommand{\meanrho}{\overline{\rho}}

{}
{}
{}

{}
{}
{}
{}
{}
{}
\newcommand{\meanAA}{\overline{\mbox{\boldmath $A$}}{}}{}
\newcommand{\meanBB}{\overline{\mbox{\boldmath $B$}}{}}{}
\newcommand{\meanUU}{\overline{\mbox{\boldmath $U$}}{}}{}
{}
{}
{}
{}
{}
{}
{}
{}
{}

\newcommand{\meanWW}{\overline{\mbox{\boldmath $W$}}{}}{}
{}
{}

\newcommand{\meanB}{\overline{B}}

\newcommand{\meanW}{\overline{W}}
\newcommand{\meanU}{\overline{U}}

\newcommand{\meanV}{\overline{V}}

\title[Turbulent magnetic helicity fluxes in solar convective zone]{Turbulent magnetic helicity fluxes in solar convective zone}

\author[N. Kleeorin and I. Rogachevskii]
{N. Kleeorin$^{1,2}$ and I. Rogachevskii$^{1,3}$ \\
$^1$Department of Mechanical Engineering, Ben-Gurion University of the Negev, P. O. B. 653, Beer-Sheva
 8410530, Israel
 \\
$^2$Institute of Continuous Media Mechanics, Korolyov str. 1, 614013 Perm, Russia
\\
$^3$Nordita, Stockholm University and KTH Royal Institute of Technology, 10691 Stockholm, Sweden}

\begin{document}


\maketitle


\begin{abstract}
Combined action of helical motions of plasma (the kinetic $\alpha$ effect) and non-uniform
(differential) rotation is a key dynamo mechanism of solar and galactic large-scale magnetic fields.
Dynamics of magnetic helicity of small-scale fields is a crucial mechanism
in a nonlinear dynamo saturation where turbulent magnetic helicity fluxes allow to avoid catastrophic
quenching of the $\alpha$ effect.
The convective zone of the Sun and solar-like stars as well as
galactic discs are the source for production of turbulent magnetic helicity fluxes.
In the framework of the mean-field approach and the spectral $\tau$ approximation,
we derive turbulent magnetic helicity fluxes
using the Coulomb gauge in a density-stratified turbulence.
The turbulent magnetic helicity fluxes include
non-gradient and gradient contributions.
The non-gradient magnetic helicity flux is proportional to
a nonlinear effective velocity (which vanishes in the absence of the density stratification)
multiplied by small-scale magnetic helicity,
while the gradient contribution describes turbulent  magnetic diffusion
of the small-scale magnetic helicity.
In addition, the turbulent magnetic helicity fluxes
contain source terms proportional to the kinetic $\alpha$ effect
or its gradients, and also contributions caused by the large-scale
shear (solar differential rotation).
We have demonstrated that the turbulent magnetic helicity fluxes due to
the kinetic $\alpha$ effect  and its radial derivative
in combination with the nonlinear magnetic diffusion
of the small-scale magnetic helicity are dominant in the solar convective zone.
\end{abstract}


\maketitle


\section{Introduction}
\label{sect1}

The large-scale solar and galactic magnetic fields are generated by a combined action
of helical turbulent motions and large-scale differential rotation due to the $\alpha\Omega$ dynamo
\citep[see, e.g.,][]{Moffatt(1978),Parker(1979),Krause(1980),Zeldovich(1983),MD2019}.
A non-zero kinetic helicity produced by a rotating density stratified convective
turbulence, causes the kinetic $\alpha$ effect.
The dynamo instability is saturated by nonlinear effects.
One of the important nonlinear effect is the feedback of the growing large-scale magnetic field
on the plasma turbulent motions, so that the turbulent transport coefficients
(the $\alpha$ effect, the effective pumping velocity and the turbulent magnetic diffusion)
depend on the mean magnetic field $\overline{\bm B}$.
The simplest nonlinear saturation mechanism of the dynamo instability
is related to the $\alpha$ quenching which prescribes the kinetic $\alpha$ effect to be a
decreasing function of the mean magnetic field strength, e.g.,
$\alpha(\overline{\bm B})=\alpha_{_{\rm K}} \, \left(1 + \overline{\bm B}^{\, 2}/
\overline{B}^{\, 2}_{\rm eq}\right)^{-1}$,
where $\alpha_{_{\rm K}} \propto  - \tau_0 \, H_{\rm u}$ is the kinetic $\alpha$ effect that is proportional to
the kinetic helicity $H_{\rm u} = \langle {\bm u} {\bf \cdot} (\bec{\nabla} {\bf \times} {\bm u}) \rangle$,
$\overline{B}^2_{\rm eq} = 4\pi \, \overline{\rho} \, \left\langle {\bm u}^2\right\rangle$
is the squared equipartition mean magnetic field, ${\bm u}$ is the turbulent velocity field,
$\tau_0$ is the turbulent time and $\overline{\rho}$ is the mean density.
This implies that the mean magnetic field strength
at which quenching becomes significant, is estimated
from the equipartition between the energy density of the mean magnetic field
and the turbulent kinetic energy density.
When applied to galactic dynamos, this picture results in robust
magnetic field models which are compatible with observations
\citep[see, e.g.,][]{Ruzmaikin(1988),SS21}.
The above-mention nonlinearity is referred as algebraic nonlinearity.

However this picture is obviously oversimplified and various
attempts to suggest a more advanced version of nonlinear
dynamo theory have been undertaken
\citep[see, e.g., reviews and books by][and references therein]{Brandenburg(2005),Ruediger(2013),RIN19,RI21}.
The quantitative theories of the algebraic nonlinearities of the
$\alpha$ effect, the turbulent magnetic diffusion and the effective pumping velocity
have been developed using the quasi-linear approach
for small fluid and magnetic Reynolds numbers
\citep{RK93,KRP94,Ruediger(2013)}
and the tau approach for large fluid and magnetic Reynolds numbers
\citep{FBC99,RK2000,RK2001,RK2004,RK2006}.

In addition to the algebraic nonlinearity, there is also a dynamic nonlinearity
caused by an evolution of magnetic helicity density of small-scale fields
during the nonlinear stage of the mean-field dynamo.
In particular, the $\alpha$ effect is the sum of the kinetic and magnetic parts,
$\alpha=\alpha_{_{\rm K}} + \alpha_{\rm m}$,
where the magnetic $\alpha$ effect, $\alpha_{\rm m} \propto \tau_0 \, H_{\rm c}/(12 \pi \, \overline{\rho})$, is proportional to
the current helicity $H_{\rm c}=\langle {\bm b} {\bf \cdot} (\bec{\nabla} {\bf \times} {\bm b}) \rangle$ of the small-scale magnetic field ${\bm b}$ \citep{PFL76}.
The dynamics of the current helicity $H_c$ is determined by the
evolution of the small-scale magnetic helicity density
$H_{\rm m} =\langle {\bm a} {\bf \cdot} {\bm b} \rangle$, where
magnetic fluctuations ${\bm b}=\bec{\nabla} {\bf \times} {\bm a}$
and ${\bm a}$ are fluctuations of magnetic vector potential.

Magnetic helicity is fundamental quantity in magnetohydrodynamics and plasma physics \citep[see, e.g.,][]{BERG99}.
In particular, the total magnetic helicity, i.e., the sum of the magnetic helicity densities of the large-scale and small-scale magnetic fields, $H_{\rm M}+H_{\rm m}$, integrated over the volume, $\int (H_{\rm M} + H_{\rm m}) \,dr^3$, is conserved for very small microscopic magnetic diffusivity $\eta$.
Here $H_{\rm M}=\overline{\bm A} {\bf \cdot} \overline{\bm B}$
is the magnetic helicity density of the large-scale field $\overline{\bm B}=\bec{\nabla} {\bf \times} \overline{\bm A}$. Signature of magnetic helicity
has been detected in many solar features, including
solar active regions \citep[see, e.g.,][and references therein]{PB14,ZKRS06,ZKRS12}.

The governing equation for small-scale magnetic helicity density $H_{\rm m}$
has been derived for an isotropic turbulence by \cite{KR82}
and for an arbitrary anisotropic turbulence
by \cite{KR99}. This equation has been
used for analytical study of solar dynamos \citep{KRR94,KRR95}
as well as for mean-field numerical modeling
of solar and galactic dynamos \citep[see, e.g.,][]{CTB97,CTB98,KMR00,KMR02,KKMR03,KMR03a,KSR16,Brandenburg(2005),
SKR06,ZKRS06,ZKRS12,DGB13,SKR18}.

As the dynamo amplifies the large-scale magnetic field,
the magnetic helicity density $H_{\rm M}$ of the large-scale field grows in time.
In particular, the evolution of the large-scale magnetic helicity density $H_{\rm M}$ is determined by
the following equation:
\begin{eqnarray}
{\partial H_{\rm M} \over \partial t} + \bec{\nabla} \cdot {\bm F}^{({\rm M})} = 2 \bec{\cal E} \cdot \overline{\bm B} - 2 \eta H_C \;,
\label{NH1}
\end{eqnarray}
where $\bec{\cal E} = \langle {\bm u}{\bf \times}{\bm b} \rangle$ is the turbulent electromotive force that determines generation and dissipation of the large-scale magnetic field, $2 \bec{\cal E} \cdot \overline{\bf B}$ is the source of $H_{\rm M}$ due to the dynamo generated large-scale magnetic field, ${\bm F}^{({\rm M})}$ is the flux of magnetic helicity density of the large-scale field that determines its transport and $H_{\rm C}=\overline{\bm B} {\bf \cdot} (\bec{\nabla} {\bf \times}  \overline{\bm B})$ is the current helicity of large-scale field.

Since the total magnetic helicity $\int (H_{\rm M} + H_{\rm m}) \,dr^3$
is conserved, the magnetic helicity density $H_{\rm m}$ of the small-scale field changes during the dynamo action, and its evolution is determined by the dynamic equation
\citep{KR82,Zeldovich(1983),KRR95,KR99}:
\begin{eqnarray}
{\partial H_{\rm m} \over \partial t} + \bec{\nabla} \cdot {\bm F}^{({\rm m})} = - 2 \bec{\cal E} \cdot \overline{\bm B} - 2 \eta H_c \;,
\label{NH2}
\end{eqnarray}
where $- 2 \bec{\cal E} \cdot \overline{\bm B}$ is the source of $H_{\rm m}$
due to the dynamo generated
large-scale magnetic field,  ${\bm F}^{({\rm m})}$ is the flux of magnetic helicity density of the small-scale field
that determines its transport
and $-2 \eta H_c$ is the dissipation rate of $H_{\rm m}$.
The source of the small-scale and large-scale magnetic helicity densities is only located in
turbulent region.

The characteristic decay time of the magnetic helicity density $H_{\rm m}$ of the small-scale field
is of the order of $T_{\rm m} =\tau_0 \, {\rm Rm}$, while the characteristic time for the decay of kinetic helicity is of the order of the turn-over time $\tau_0=\ell_0 / u_0$ of turbulent eddies in the
integral turbulence scale $\ell_0$, where ${\rm Rm} = \ell_0 \, u_0 /\eta$ is the magnetic Reynolds number.
The current helicity $H_{\rm c}$ of the small-scale field is not a conserved quantity, and the characteristic decay time of $H_{\rm c}$ varies from a short timescale $\tau_0$ to much larger timescales. On the other hand, the characteristic decay times of the current helicity of large-scale field, $H_{\rm C}$, and of the large-scale magnetic helicity $H_{\rm M}$ are of the
order of the turbulent diffusion time. For weakly inhomogeneous
turbulence the current helicity  density $H_{\rm c}$ of the small-scale field is proportional
to the small-scale magnetic helicity  density $H_{\rm m}$ \citep{KR99}.

Using the steady-state solution of Eq.~(\ref{NH2}) with a zero turbulent
flux ${\bm F}^{({\rm m})}=0$ of magnetic helicity density of small-scale field
and a zero current helicity of large-scale field, $H_{\rm C}$,
it has been concluded that the critical mean magnetic field strength,
$\overline{B}_{\rm cr}$, at which the dynamic $\alpha$ quenching becomes significant, in fact  is much lower than
the equipartition value, e.g. $\overline{B}_{\rm cr} = \overline{B}_{\rm eq} \, {\rm Rm}^{-1/2}$
\citep{VC92,GD94}.
In astrophysics, e.g., in galactic disks and in the convective zone of the sun, magnetic
Reynolds numbers are very large.
Therefore, for large magnetic Reynolds numbers
the dynamo action should saturate at a magnetic field strength that is much
lower than the equipartition value. This effect is referred as to a catastrophic
quenching of the $\alpha$ effect \citep{VC92,GD94}.
On the other hand, the  observed large-scale field strengths in spiral galaxies
is of the order of the equipartition value  \citep[see, e.g.,][]{Ruzmaikin(1988),SS21},
and the observed solar and stellar magnetic fields are much larger than $\overline{B}_{\rm cr}$
\citep[see, e.g.,][]{Moffatt(1978),Parker(1979),Krause(1980),Zeldovich(1983)}.

The evolution of magnetic helicity appears however to be a more
complicated process than can simply be described by a balance of
magnetic helicity in a given volume. It is necessary to take
into account fluxes of magnetic helicity \citep{KMR00}. This implies that the
turbulent transport of magnetic helicity through the boundaries
(the open boundary conditions in simulations) should be taken into account  \citep{BF00}.
Different forms of magnetic helicity fluxes have been suggested in various studies
\citep{CTB97,CTB98,KR99,KMR00,KMR02,VC01,SB04,Brandenburg(2005)}.
Turbulent fluxes of small-scale magnetic helicity have been measured in numerical simulations
\citep{KKB10,MCB10,HB10,HB11,HB12,DGB13},
and in solar observations \citep{CWQ01,PDB05,PB14,HB18}.

Taking into account turbulent fluxes of the small-scale magnetic helicity,
it has been shown by numerical simulations that a nonlinear galactic dynamo
governed by a dynamic equation for the
magnetic helicity density $H_{\rm m}$ of small-scale field
saturates at a mean magnetic field comparable with the
equipartition magnetic field
\citep[see, e.g.,][]{KMR00,KMR02,KKMR03,KMR03a,BB02,Brandenburg(2005),SSS06,DGB13}.
Numerical simulations demonstrate that the dynamics of the small-scale magnetic helicity
in the presence of the turbulent magnetic helicity fluxes
play a crucial role in the solar dynamo as well
\citep[see, e.g.,][]{KKMR03,KSR16,KSR20,SKR06,ZKRS06,ZKRS12,KKB10,GCB10,HB12,DGB13,SKR18,RIN21}.

Due to very important role of the turbulent magnetic helicity fluxes
in nonlinear dynamos,
in the present study we perform a rigorous derivation of these fluxes
applying the mean-field theory, adopting the Coulomb gauge and
considering a strongly density-stratified turbulence.
We show that the turbulent magnetic helicity fluxes contain
non-gradient and gradient contributions.
The non-gradient magnetic helicity fluxes are product of
a nonlinear effective velocity and small-scale magnetic helicity.
The gradient contributions determine a nonlinear magnetic diffusion
of the small-scale magnetic helicity.
We also demonstrate that the turbulent magnetic helicity fluxes
include source terms proportional to the kinetic $\alpha$ effect
or its gradients.
In the present study we do not consider an algebraic quenching of
the turbulent magnetic helicity fluxes that is a subject of a separate study.

This paper is organized as follows.
In Section~\ref{sect2}, we derive equation for the magnetic helicity of small-scale fields
which includes divergence of the turbulent magnetic helicity flux.
In Section~\ref{sect3} we discuss the results of calculations
of the turbulent flux of magnetic helicity of the small-scale fields.
In addition, we obtain a general form of turbulent flux of the magnetic helicity
using symmetry arguments.
In Section~\ref{sect4}, we consider the turbulent magnetic helicity
flux in the solar convective zone.
Finally, in Section~\ref{sect5}, we discuss our results and draw conclusions.
In Appendixes~\ref{Appendix A} and~\ref{Appendix B} we discuss approximations and procedure
of the derivation of turbulent flux of magnetic helicity.
In Appendix~C we determine the effect of large-scale shear on turbulent flux of the magnetic helicity.
Applying the method described in Appendixes~\ref{Appendix A}--\ref{Appendix C},
we determine various contributions  to the turbulent flux of the small-scale magnetic helicity
in Appendix~\ref{Appendix D}.
In particular, we present the general form of turbulent transport coefficients
entering in the turbulent flux of the small-scale magnetic helicity.
For better understanding of the physics related to various contributions
to the turbulent flux of the small-scale magnetic helicity, in Appendix~\ref{Appendix E}
we consider a more simple case with a large-scale linear velocity shear
and present turbulent transport coefficients in the Cartesian coordinates.

\section{Equation for the magnetic helicity}
\label{sect2}

In this Section, we derive an equation for the small-scale magnetic helicity.
The induction equation for fluctuations of magnetic field ${\bm b}$
reads
\begin{eqnarray}
&& {\partial {\bm b} \over \partial t} = {\bm \nabla} \times
\Big[\meanUU \times {\bm b} + {\bm u} \times \meanBB
+ {\bm u} \times {\bm b} - \langle{\bm u} \times {\bm b}\rangle
\nonumber\\
&& \qquad - \eta \bec{\bf \nabla} \times {\bm b}\Big],
\label{PP1}
\end{eqnarray}
where in the framework of the mean-field approach, we separate
magnetic and velocity fields into mean and fluctuations,
${\bm B} = \meanBB + {\bm b}$ and $\meanBB = \langle
{\bm B} \rangle $ is the mean magnetic field,
${\bm U} = \meanUU + {\bm u}$, and $\meanUU = \langle
{\bm U} \rangle$ is the mean fluid velocity
describing, e.g., the differential rotation, $\eta$ is the
magnetic diffusion due to electrical conductivity of fluid.
The equation for magnetic fluctuations is obtained by subtracting induction equation for the
the mean magnetic field $\meanBB$ from that for the total
field ${\bm B}(t,{\bm x})$.
The equation for fluctuations of the vector potential ${\bm a}$ follows
from induction equation (\ref{PP1})
\begin{eqnarray}
&& {\partial {\bm a} \over \partial t} = \meanUU \times {\bm b} + {\bm u} \times \meanBB
+ {\bm u} \times {\bm b} - \langle{\bm u} \times {\bm b}\rangle
\nonumber\\
&& \qquad - \eta \bec{\bf \nabla} \times {\bm b} + {\bm \nabla} \phi,
\label{PP2}
\end{eqnarray}
where ${\bm B} = \bec{\bf \nabla} \times {\bm A}$ and
${\bm A} = \meanAA + {\bm a} ,$ and $\meanAA = \langle
{\bm A} \rangle $ is the mean vector potential,
${\bm b} = \bec{\bf \nabla} \times {\bm a}$ and
$\phi$ are fluctuations of the scalar potential.
We multiply Eq.~(\ref{PP1}) by ${\bm a}$ and Eq.~(\ref{PP2}) by ${\bm b}$, add them and
average over an ensemble of turbulent fields. This yields an equation
for the magnetic helicity $H_{\rm m} = \langle {\bm a}({\bf x}) \cdot
{\bm b}({\bf x}) \rangle$ of the small-scale fields as
\begin{eqnarray}
{\partial H_{\rm m} \over \partial t} = - 2\bec{\cal E}  \cdot \meanBB - 2 \eta \langle {\bm b} \cdot ({\bm \nabla} \times {\bm b}) \rangle - {\bm \nabla} \cdot {\bm F}^{({\rm m})} ,
\label{PP3}
\end{eqnarray}
where $\bec{\cal E} =\langle {\bm u} \times {\bm b} \rangle$
is the turbulent electromotive force,
and the turbulent flux of magnetic helicity ${\bm F}^{({\rm m})}$ of the small-scale fields is
given by
\begin{eqnarray}
&& {\bm F}^{({\rm m})} = \meanUU \, H_{\rm m} - \left\langle {\bm b} \,({\bm a} \cdot
\meanUU) \right\rangle + \left\langle {\bm u} \,({\bm a} \cdot
\meanBB) \right\rangle - \meanBB \left\langle {\bm a} \cdot
{\bm u} \right\rangle
\nonumber\\
&& \quad - \eta \left\langle {\bm a} \times ({\bm \nabla}
\times{\bm b}) \right\rangle +\left\langle {\bm a} \times ({\bm u}
\times{\bm b}) \right\rangle - \left\langle {\bm b} \, \phi \right\rangle .
\label{PP4}
\end{eqnarray}
Using the Coulomb gauge ${\bm \nabla}\cdot{\bm a}=0$,
we obtain that ${\bm \nabla}\times{\bm b} = - \Delta {\bm a}$
and  ${\bm a} = - \Delta^{-1} \, {\bm \nabla}\times{\bm b}$.
The Coulomb gauge also allows us
to find fluctuations of the scalar potential $\phi$.
Indeed, equation for ${\bm \nabla}\cdot{\bm a}$
which follows from Eq.~(\ref{PP2}),
yields expression for fluctuations of the scalar potential $\phi$, so that
the correlation function $\left\langle b_i \, \phi \right\rangle$ reads
\begin{eqnarray}
&&\left\langle b_i \, \phi \right\rangle = \left\langle b_i \,a_j \right\rangle\, \meanU_j
- \left\langle b_i \,\Delta^{-1} \,
({\bm \nabla}\times{\bm u})_j \right\rangle\, \meanB_j
\nonumber\\
&& \quad
- \left\langle b_i \,\Delta^{-1} \, b_j \right\rangle \, \meanW_j + \left\langle b_i \,\Delta^{-1} \, u_j \right\rangle \,
({\bm \nabla}\times\meanBB)_j
\nonumber\\
&& \quad - \left\langle b_i \,\Delta^{-1} \, {\bm \nabla}\cdot({\bm u}
\times{\bm b}) \right\rangle .
\label{PP5}
\end{eqnarray}
where $\meanWW = \bec{\bf \nabla} \times \meanUU$ is the mean vorticity and $\left\langle b_i \,a_j \right\rangle = - \left\langle b_i \,\Delta^{-1} \,
({\bm \nabla}\times{\bm b})_j \right\rangle$.
Equations~(\ref{PP4})--(\ref{PP5}) yield the turbulent flux of
magnetic helicity ${\bm F}^{({\rm m})}$ of the small-scale fields as
\begin{eqnarray}
&& F_i^{({\rm m})} = \meanU_i \, H_{\rm m} + \meanW_j  \,\left\langle b_i \,\Delta^{-1} \, b_j \right\rangle
+ \meanB_j \, \left\langle u_i \, a_j\right\rangle
\nonumber\\
&& \quad - \meanB_i \, \left\langle u_j \, a_j\right\rangle + \meanB_j \, \left\langle b_i \,\Delta^{-1} \,
({\bm \nabla}\times{\bm u})_j \right\rangle
 + F_i^{(\eta)}
 \nonumber\\
&& \quad
 - ({\bm \nabla}\times\meanBB)_j \,\left\langle b_i \,\Delta^{-1} \, u_j \right\rangle + F_i^{({\rm III})},
\label{PP6}
\end{eqnarray}
where $\left\langle u_i \, a_j\right\rangle = - \left\langle u_i \, \,\Delta^{-1} \,({\bm \nabla}\times{\bm b})_j\right\rangle$,
${\bm F}^{(\eta)} = - \eta \left\langle {\bm a} \times ({\bm \nabla}
\times{\bm b}) \right\rangle$ is the flux caused by the microscopic magnetic diffusion $\eta$
and ${\bm F}^{({\rm III})}$ is the flux that is determined by the third-order moments, and it is given by
\begin{eqnarray}
&& {\bm F}^{({\rm III})} = \left\langle {\bm b} \,\Delta^{-1} \, {\bm \nabla}\cdot({\bm u}
\times{\bm b}) \right\rangle + \left\langle {\bm a} \times ({\bm u}
\times{\bm b}) \right\rangle .
\label{PP7}
\end{eqnarray}
Equations~(\ref{PP3})--(\ref{PP7}) are exact equations.
Note that only in the Coulomb gauge, the scalar potential $\phi$
is described by the stationary equation.
For all other gauge conditions, the scalar potential $\phi$
is determined by a non-stationary equation.
Also for the Coulomb gauge the relation between the magnetic $\alpha$
effect and small-scale magnetic helicity is most simple.

\section{General form of turbulent flux of the magnetic helicity}
\label{sect3}

In this Section we discuss the results of calculations
of the turbulent flux of magnetic helicity of the small-scale fields.
General form of turbulent flux ${\bm F}^{({\rm m})}$ of the magnetic helicity can be obtained
from symmetry reasoning. Indeed, the turbulent flux ${\bm F}^{({\rm m})}$ is the pseudo-vector
which should contain two pseudo-scalars:
the magnetic helicity, $H_{\rm m}$, and the kinetic $\alpha$ effect, $\alpha_{_{\rm K}}$, and their first spatial derivatives.
In addition, the contributions $F_i^{({\rm S}0)}$ to the turbulent magnetic helicity flux
caused by the large-scale shear (differential rotation) should contain the pseudo-vector
$\meanWW={\bm \nabla} \times \meanUU$,
where $\meanUU={\bm \delta\Omega} \times {\bm r}$ is the large-scale velocity describing
the differential rotation ${\bm \delta\Omega}$.

All turbulent transport coefficients entering in  the turbulent flux ${\bm F}^{({\rm m})}$
of magnetic helicity of the small-scale fields
should be quadratic in the large-scale magnetic field $\meanBB$,
i.e., they should be proportional to $\meanB^2$ or $\meanV_{\rm A}^2=\meanB^2/(4 \pi \meanrho)$,
where $\meanrho$ is the mean plasma density and $\meanV_{\rm A}$ is the mean Alfv\'{e}n speed.
On the other hand, the turbulent flux ${\bm F}^{({\rm m})}$ of the magnetic helicity should vanish
in the absence of turbulence. This implies that all turbulent transport coefficients
entering in the turbulent flux ${\bm F}^{({\rm m})}$  should be proportional
to turbulent correlation time $\tau_0$ or turbulent integral scale $\ell_0$.
Some of the turbulent transport coefficients are caused by the plasma density stratification, i.e., they are
proportional to ${\bm \lambda} = - {\bm \nabla} \ln \meanrho$.

Using the theoretical approach based on the spectral $\tau$ approximation
which is valid for large fluid and magnetic Reynolds numbers,
and the multi-scale approach,
we obtain the turbulent flux of the small-scale magnetic helicity as
\begin{eqnarray}
&& F_i^{({\rm m})} = \left(\meanU_i + V_i^{({\rm H})}\right) \, H_{\rm m} - D_{ij}^{({\rm H})} \, \nabla_j H_{\rm m}
+ N_i^{(\alpha)} \, \alpha_{_{\rm K}}
 \nonumber\\
&& \quad+ M_{ij}^{(\alpha)} \, \nabla_j \alpha_{_{\rm K}} +
F_i^{({\rm S}0)},
\label{RR1}
\end{eqnarray}
where $\alpha_{_{\rm K}}=- \tau_0 \, H_{\rm u} /3$ is the kinetic $\alpha$ effect.
Details of the derivation of  Eq.~(\ref{RR1}) are described in Appendixes~\ref{Appendix A}--\ref{Appendix C}.
The general form of the turbulent transport coefficients entering in  the turbulent flux~(\ref{RR1})
of magnetic helicity of the small-scale fields is given by Eqs.~(\ref{RR20})--(\ref{PPFF25})
in Appendix~\ref{Appendix D}.
These turbulent transport coefficients of the turbulent magnetic helicity flux in spherical coordinates
are given in the next section and in the Cartesian coordinates are discussed in Appendix~\ref{Appendix E}.

The turbulent flux of the small-scale magnetic helicity
includes the non-gradient and gradient contributions.
The non-gradient contribution to the turbulent  flux of magnetic helicity is proportional to
the sum of the mean velocity $\meanUU={\bm \delta\Omega} \times {\bm r}$ and the turbulent pumping velocity ${\bm V}^{({\rm H})}$
which is multiplied by small-scale magnetic helicity $H_{\rm m}$,
while the gradient contribution $- D_{ij}^{({\rm H})} \, \nabla_j H_{\rm m}$
describe the turbulent magnetic diffusion of the small-scale magnetic helicity.
The effective pumping velocity of the small-scale magnetic helicity
${\bm V}^{({\rm H})}$ vanishes in the absence of the density stratification.
In addition, the turbulent magnetic helicity flux
contains the source term ${\bm N}^{(\alpha)} \, \alpha_{_{\rm K}}$
proportional to the kinetic $\alpha$ effect, and
the source term $- M_{ij}^{(\alpha)} \, \nabla_j \alpha_{_{\rm K}}$
proportional to the gradient $\nabla_j \alpha_{_{\rm K}}$
of the kinetic $\alpha$ effect.
The turbulent magnetic helicity flux
also have contributions caused by the large-scale
shear (differential rotation) in the turbulent flow.

We assume that the turbulent flux of the magnetic helicity ${\bm F}^{({\rm III})}$ containing the third-order moments [see equation (\ref{PP7})], is determined using the turbulent diffusion approximation as
${\bm F}^{({\rm III})}=-D_{T}^{({\rm H})} \, {\bm \nabla} H_{\rm m}$.
The contribution to the turbulent magnetic helicity flux,  $-D_{T}^{({\rm H})} \, {\bm \nabla} H_{\rm m}$,
caused by the turbulent diffusion, has been used in mean-field numerical simulations by \cite{CTB97,CTB98,KMR02,KMR03a}.

The turbulent diffusion of the small-scale magnetic helicity
can be interpreted as follows. The random
flows existing in the interstellar medium consist of a combination
of small-scale motions, which are affected by magnetic
forces (tangling turbulence) resulting in a steady-state of the dynamo, and a background micro-turbulence
which is supported by a strong random driver (e.g., supernovae
explosions which can be considered as independent
of the galactic magnetic field). The large-scale magnetic field
is smoothed over both kinds of turbulent fluctuations, while
the small-scale magnetic field is smoothed over micro-turbulent
fluctuations only. It is the smoothing over the micro-turbulent
fluctuations that gives the coefficient $D_{T}^{({\rm H})}
= C_D \eta_{_{T}}$  with a free dimensionless constant $C_D  \sim 0.1$.
Here $\eta_{_{T}}$ is the turbulent diffusion coefficient of the mean magnetic field.

The magnetic helicity flux ${\bm F}^{(\eta)} = - \eta \left\langle {\bm a} \times ({\bm \nabla}
\times{\bm b}) \right\rangle$ due to the microscopic magnetic diffusion $\eta$ is given by
${\bm F}^{(\eta)} = - {1 \over 3} \eta {\bm \nabla} H_{\rm m}$.
This flux in astrophysical systems is very small and neglected here.

\section{Turbulent magnetic helicity flux in the solar convective zone}
\label{sect4}

In this Section we discuss the results of calculations of the turbulent magnetic helicity
flux in the solar convective zone, where we use spherical coordinates $(r, \vartheta, \varphi)$.
The radial turbulent flux of the small-scale magnetic helicity is given by
\begin{eqnarray}
&& F_r^{({\rm m})} = V_r^{({\rm H})} \, H_{\rm m} - D_{rj}^{({\rm H})} \, \nabla_j H_{\rm m}
+ N_r^{(\alpha)} \, \alpha_{_{\rm K}}
 \nonumber\\
&& \quad+ M_{rj}^{(\alpha)} \, \nabla_j \alpha_{_{\rm K}} +
F_r^{({\rm S}0)} .
\label{SRR1}
\end{eqnarray}
The general forms of the turbulent transport coefficients entering in  the turbulent flux ${\bm F}^{({\rm m})}$
of magnetic helicity of the small-scale fields are given by Eqs.~(\ref{RR20})--(\ref{PPFF25}) in Appendix~\ref{Appendix D}.
In view of applications  to the solar convective zone,
the turbulent transport coefficients of the turbulent magnetic helicity flux in spherical coordinates are specified below:
\begin{eqnarray}
V_r^{({\rm H})} = - {1 \over 15} \tau_0 \, \meanV_{\rm A}^2 \, \lambda \, \biggl[1 +
7 \beta_r^2 - {173 \over 14} \, \sin \vartheta \, \tau_0 \, \delta\Omega \, \beta_r \beta_\varphi\biggr] ,
\label{RR2}
\end{eqnarray}

\begin{eqnarray}
D_{rr}^{({\rm H})} = D_{T}^{({\rm H})} + {1 \over 30} \tau_0 \, \meanV_{\rm A}^2 \, \left(5 - 4  \beta_r^2\right) ,
\label{RR4}
\end{eqnarray}

\begin{eqnarray}
D_{r\vartheta}^{({\rm H})} =  {2(80 + 17 q) \over 105} \tau_0^2 \, \meanV_{\rm A}^2  \,\delta\Omega \,\beta_r \, \beta_\varphi \,
\cos \vartheta  ,
\label{RRR4}
\end{eqnarray}

\begin{eqnarray}
N_r^{(\alpha)} &=& - {1 \over 10} \ell_0^2 \, \meanB^2 \, \lambda \, \biggl[1 + {7q-2\over q} \, \beta_r^2
 \nonumber\\
&& - {216(q-1)\over 7(3q-1)} \, \tau_0 \, \delta\Omega\,\beta_r \,\beta_\varphi \, \sin \vartheta \biggr],
\label{RR7}
\end{eqnarray}

\begin{eqnarray}
M_{rr}^{(\alpha)} &=& {2q -1\over 20 q} \ell_0^2 \, \meanB^2 \, \biggl[1 +
{20 q -23 \over 2q-1} \, \beta_r^2
\nonumber\\
&&- {32 q (q-1) \over (2q-1) \, (3q-1)} \, \tau_0 \, \delta\Omega \,\beta_r \,\beta_\varphi \, \sin \vartheta \biggr] ,
\label{RR9}
\end{eqnarray}

\begin{eqnarray}
M_{r\vartheta}^{(\alpha)} &=& {8(q-1) \over 3q-1} \ell_0^2 \, \meanB^2 \, \tau_0 \,\delta\Omega  \, \beta_r \,\beta_\varphi
\,  \cos \vartheta   ,
\label{RR10}
\end{eqnarray}

\begin{eqnarray}
F_r^{({\rm S}0)} &=& - {2 \over 9}  \, \delta\Omega  \,  \cos \vartheta \, \biggl\{4
 \, \ell_0^2 \, \meanB_r^2 + \biggl[{\meanV_{\rm A}^2  \over \left\langle {\bm u}^2 \right\rangle} \, \Big(1- {3 \over 11} \beta_r^2\Big)
 \nonumber\\
&&
 +{3(q-1) \over q+1} \biggr]\, \ell_b^2 \, \left\langle {\bm b}^2 \right \rangle \biggr\} ,
\label{RR12}
\end{eqnarray}
where ${\bm \beta}=\meanBB/\meanB$ is the unit vector along the mean magnetic field,
$\meanUU= \delta\Omega \, r \, \sin\vartheta\, {\bm e}_\varphi$ is the mean velocity caused by the differential rotation $\delta\Omega = \Omega(r,\vartheta) - \Omega(r=R_\odot,\vartheta)$.
Here $\Omega(r=R_\odot,\vartheta) =\Omega_0 (1 - C_2 \cos^2 \vartheta - C_4 \cos^4 \vartheta)$
with $\Omega_0=2.83 \times 10^{-6}$ s$^{-1}$, $C_2=0.121$ and $C_4=0.173$ \citep{LH82},
$R_\odot$ is the solar radius, ${\bm \lambda}= \lambda \, {\bm e}_r$,
$\ell_b$ is the energy containing scale of magnetic fluctuations with a zero mean magnetic field
and $q$ is the exponent in the spectrum of the turbulent kinetic energy
(the exponent $q=5/3$ corresponds to the Kolmogorov spectrum of the turbulent kinetic energy).

In derivation of Eqs.~(\ref{RR2})--(\ref{RR12}), we take into account that
for weakly inhomogeneous turbulence
$H_{\rm c} \approx H_{\rm m}/\ell_0^2$, and we neglect small terms  $\sim {\rm O}[\ell_0^2/L_{\rm m}^2]$ with $L_{\rm m}$ being characteristic scale of spatial variations of $H_{\rm m}$.
We neglect also small contributions proportional to spatial derivatives of the mean magnetic field,
and spatial derivatives of $\left\langle {\bm u}^2 \right\rangle$ and $\delta\Omega$.

\begin{figure}
\centering
\includegraphics[width=7.5cm]{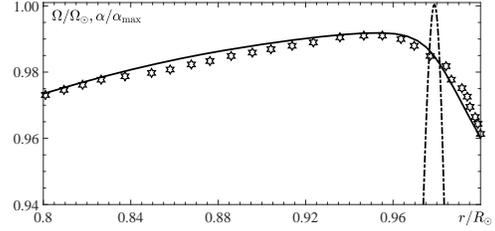}
\caption{\label{Fig1} The theoretical radial profiles of the total angular
velocity $\Omega(r)/\Omega_\odot$ (solid) that includes the uniform
and differential rotation specified for the latitude $\phi_\ast=30^\circ$
and the normalized kinetic $\alpha$ effect, $\alpha_{_{\rm K}} / \alpha_{\rm max}$ (dashed).
The theoretical profile of the total angular velocity is compared with the radial profile of the solar angular velocity
obtained from the helioseismology observational data (stars)  specified for the latitude $\phi_\ast=30^\circ$ and normalized by the solar rotation frequency $\Omega_\odot(\phi_\ast=0)$ at the equator (Kosovichev et al. 1997), where $R_\odot$ is the solar radius.
The profile $ \alpha_{_{\rm K}}(r) \equiv \alpha^{^{({\rm K})}}_{ \varphi \varphi}$ is given by Eq.~ (22) derived by Kleeorin \& Rogachevskii (2003), and $\Omega(r)/\Omega_\odot$ is given by Eq.~ (3.14) derived by Rogachevskii  \& Kleeorin (2018).}
\end{figure}

\begin{figure}
\centering
\includegraphics[width=7.5cm]{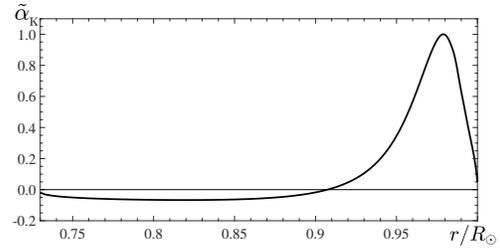}
\caption{\label{Fig2} The radial profile of the normalized kinetic $\alpha$ effect, $\tilde \alpha_{_{\rm K}} = \alpha_{_{\rm K}} / \alpha_{\rm max}$, specified for the latitude $\phi_\ast=30^\circ$ and given by Eq.~ (22) derived by Kleeorin \& Rogachevskii (2003).}
\end{figure}

Let us discuss the obtained results. For illustration, in Fig.~\ref{Fig1} we show the radial profile of the total angular velocity $\Omega(r)/\Omega_\odot$ in the solar convective zone that includes the uniform and differential rotation specified for the latitude $\phi_\ast=30^\circ$ .
The theoretical profile (solid line) of the total angular velocity \citep{RK18}
is compared with the radial profile of the solar angular velocity (stars)
obtained from the helioseismology observational data  \citep{kosovichev1997} specified for the latitude $\phi=30^\circ$
and normalized by the solar rotation frequency $\Omega_\odot(\phi_\ast=0)$ at the equator, where
$\Omega/\Omega_\odot$ is given by Eq.~ (3.14) derived by  \cite{RK18}.
In Figs.~\ref{Fig1}--\ref{Fig2} we also show the radial profile of the kinetic $\alpha$ effect,
$\alpha_{_{\rm K}} / \alpha_{\rm max}$ which is specified for the latitude $\phi=30^\circ$ and
given by Eq.~ (22) derived by  \cite{KR03}.

In the upper part of the solar convective zone for the latitude $\phi_\ast>0$ (the Northern Hemisphere),
the kinetic $\alpha$ effect is positive, $\alpha_{_{\rm K}}> 0$ (see Fig.~\ref{Fig2}).
On the other hand, the magnetic $\alpha$ effect in this region is negative,
i.e., $\alpha_{_{\rm M}} = \tau_0 \, H_{\rm c} /(4 \pi \meanrho) < 0$.
This implies that the current helicity $H_{\rm c}< 0$ as well as the magnetic helicity $H_{\rm m}< 0$ are negative
the Northern Hemisphere.
Here for simplicity, we choose the radial profile of the poloidal and toroidal field as
$\overline{B}_r=\overline{B}_{r0} \, \sin [\pi (r-0.73 R_\odot)/(0.6 R_\odot)]$ and
$\overline{B}_\varphi=\overline{B}_{\varphi0} \, \cos [\pi (r-0.73 R_\odot)/(0.6 R_\odot)]$, where
$\overline{B}_{r0}$ is the surface mean magnetic field measured in Gauss.
To avoid catastrophic quenching, the radial component of the turbulent flux of the small-scale magnetic helicity
$F_r^{({\rm m})}< 0$ should be negative for the Northern Hemisphere.

\begin{figure}
\centering
\includegraphics[width=7.5cm]{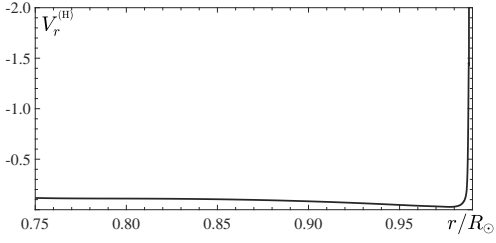}
\caption{\label{Fig3} The radial profile of the effective pumping velocity  $V_r^{({\rm H})}$ of the small-scale magnetic helicity
given by Eq.~(\ref{RR2}), and measured  in m s$^{-1}$.}
\end{figure}

\begin{figure}
\centering
\includegraphics[width=7.5cm]{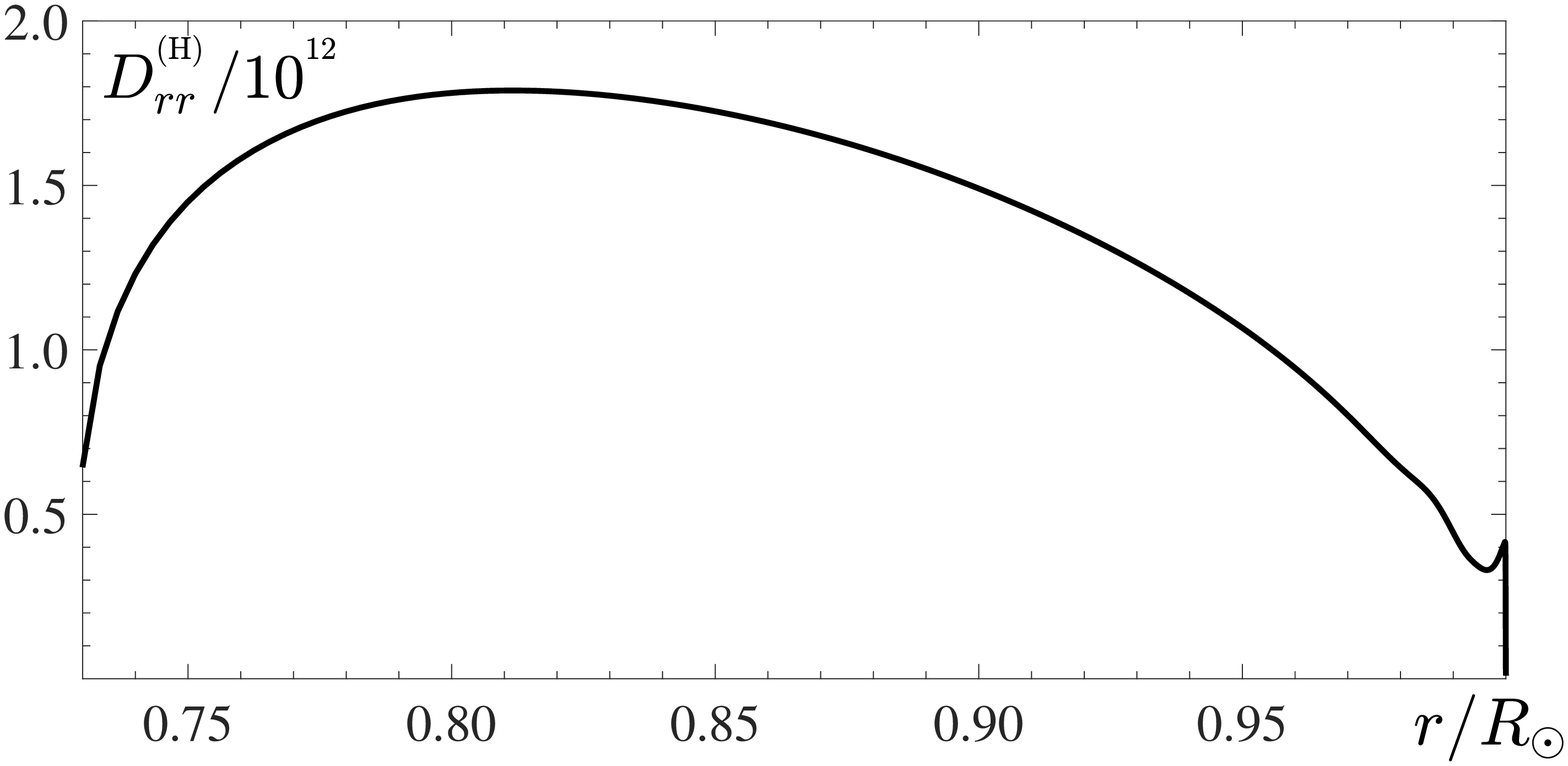}
\caption{\label{Fig4} The radial profile of turbulent diffusion $D_{rr}^{({\rm H})}(r)$ of the small-scale magnetic helicity
given by Eq.~(\ref{RR4}) and measured in cm$^2$ s$^{-1}$.}
\end{figure}

\begin{figure}
\centering
\includegraphics[width=7.5cm]{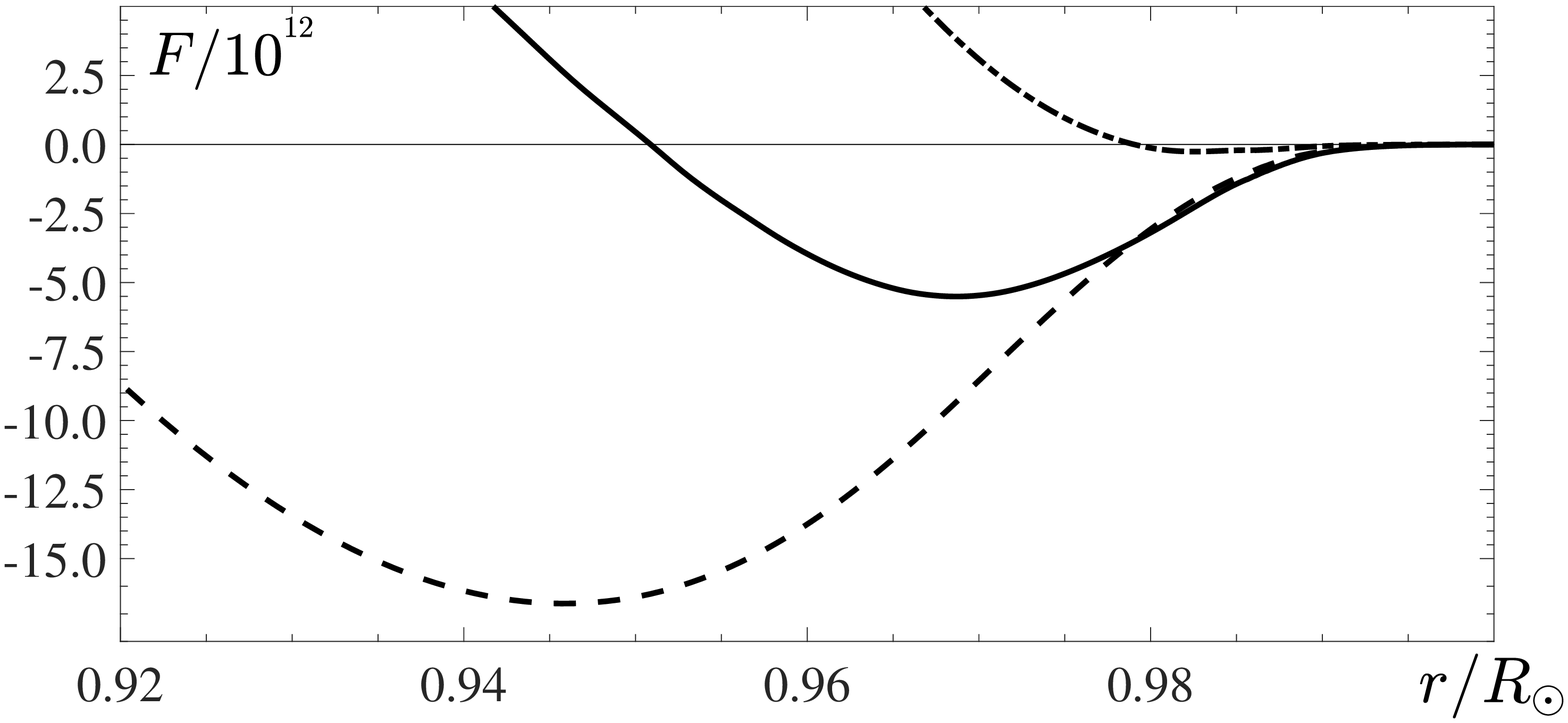}
\caption{\label{Fig5} The radial profile of the turbulent magnetic helicity fluxes caused by
the source terms $F_{1}^{(\alpha)} =N_r^{(\alpha)} \, \alpha_{_{\rm K}}$ (dashed) and
$F_{2}^{(\alpha)} =M_{rr}^{(\alpha)} \, \nabla_r \alpha_{_{\rm K}}$ (dashed-dotted)
which are proportional to the kinetic $\alpha$ effect and its radial derivative,
as well as their sum $F_r^{(\alpha)} = N_r^{(\alpha)} \, \alpha_{_{\rm K}}
+ M_{rr}^{(\alpha)} \, \nabla_r \alpha_{_{\rm K}}$ (solid), where $N_r^{(\alpha)}$ and $M_{rr}^{(\alpha)}$ are
given by Eqs.~(\ref{RR7}) and~(\ref{RR9}), respectively.
The fluxes are specified  for the latitude $\phi_\ast=30^\circ$ and measured  in G$^2$ cm$^2$ s$^{-1}$.}
\end{figure}

\begin{figure}
\centering
\includegraphics[width=7.5cm]{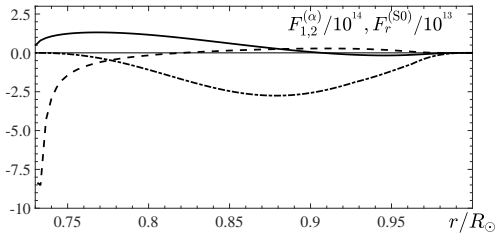}
\caption{\label{Fig6} The radial profiles of the turbulent magnetic helicity fluxes caused by
the source terms $F_{1}^{(\alpha)} =N_r^{(\alpha)} \, \alpha_{_{\rm K}}$ (solid),  $F_{2}^{(\alpha)} =M_{rr}^{(\alpha)} \, \nabla_r \alpha_{_{\rm K}}$ (dashed)
and the  contribution $F_r^{({\rm S}0)}$ (dashed-dotted) to the turbulent magnetic helicity flux caused by the large-scale shear (differential rotation) , where $N_r^{(\alpha)}$, $M_{rr}^{(\alpha)}$ and $F_r^{({\rm S}0)}$ are
given by Eqs.~(\ref{RR7}), (\ref{RR9}) and~(\ref{RR12}), respectively. The fluxes are specified  for the latitude $\phi_\ast=30^\circ$ and measured  in G$^2$ cm$^2$ s$^{-1}$.}
\end{figure}

\begin{figure}
\centering
\includegraphics[width=7.5cm]{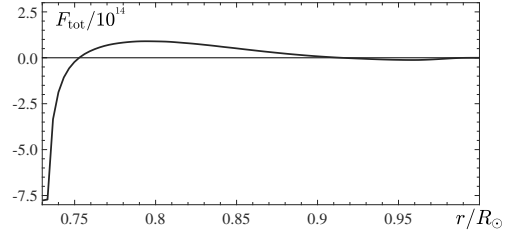}
\caption{\label{Fig7} The radial profile of the total source flux $F_{\rm tot} = N_r^{(\alpha)} \, \alpha_{_{\rm K}} + M_{rr}^{(\alpha)} \, \nabla_r \alpha_{_{\rm K}} + F_r^{({\rm S}0)}$ of the magnetic helicity that is independent of the magnetic helicity and its radial derivative. Here the flux is measured  in G$^2$ cm$^2$ s$^{-1}$.}
\end{figure}

In Figs.~\ref{Fig3} and~\ref{Fig4} we show the radial profiles of
the effective pumping velocity $V_r^{({\rm H})}(r)$
and turbulent diffusion $D_{rr}^{({\rm H})}(r)$ of the small-scale magnetic helicity.
In Figs.~\ref{Fig5} and~\ref{Fig6} we plot the radial profiles of
the turbulent magnetic helicity fluxes caused by
the source terms $F_{1}^{(\alpha)}(r) =N_r^{(\alpha)} \, \alpha_{_{\rm K}}$
and $F_{2}^{(\alpha)}(r) =M_{rr}^{(\alpha)} \, \nabla_r \alpha_{_{\rm K}}$,
which are proportional to the kinetic $\alpha$ effect and its radial derivative,
as well as their sum
$F_r^{(\alpha)}(r) = N_r^{(\alpha)} \, \alpha_{_{\rm K}} + M_{rr}^{(\alpha)} \, \nabla_r \alpha_{_{\rm K}}$.
In Fig.~\ref{Fig6} we also show
the  contribution ${\bm F}^{({\rm S}0)}(r)$ to the turbulent magnetic helicity flux caused by the large-scale shear (differential rotation).
Finally, in Fig.~\ref{Fig7} we plot the radial profile of the total source flux of the magnetic helicity $F_{\rm tot}(r) = N_r^{(\alpha)} \, \alpha_{_{\rm K}} + M_{rr}^{(\alpha)} \, \nabla_r \alpha_{_{\rm K}} + F_r^{({\rm S}0)}$ that is independent of the magnetic helicity and its radial derivative.

As follows from Figs.~\ref{Fig3}--\ref{Fig7} as well as Eqs.~(\ref{SRR1})--(\ref{RR12}),
the negative contribution to the turbulent magnetic helicity flux $F_r^{({\rm m})}$
in the range of the generation of the mean magnetic field,
is due to the source flux $F_r^{(\alpha)} = N_r^{(\alpha)} \, \alpha_{_{\rm K}} + M_{rr}^{(\alpha)} \, \nabla_r \alpha_{_{\rm K}}$,  and
the  contribution ${\bm F}^{({\rm S}0)}$ to the turbulent magnetic helicity flux caused by the large-scale shear (differential rotation).
Here we take into account that $\delta\Omega > 0$ at $0.8< r/R_\odot<1$ (see Fig.~\ref{Fig1}), where
the differential rotation $\delta\Omega = \Omega(r) - \Omega(r=R_\odot)$.

\begin{figure}
    \centering
\includegraphics[width=7.5cm]{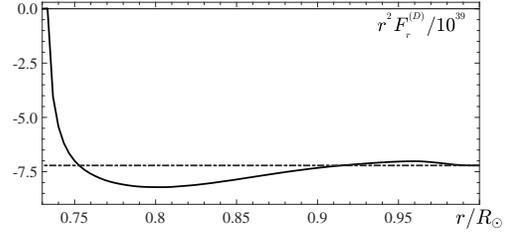}
\caption{\label{Fig8} Turbulent diffusion flux $r^2 F_r^{(D)}$ (solid line)
and the flux $r^2\, [F_r^{(D)}(r) + F_{\rm tot}(r)]$ (dashed-dotted line) of magnetic helicity
per unit solid angle which are measured in Mx$^{2}$ \, h$^{-1}$.}
\end{figure}

The small-scale magnetic helicity is not accumulated inside
the solar convective zone due to turbulent magnetic diffusion flux, $F_r^{(D)}$.
In Fig.~\ref{Fig8} we show the turbulent diffusion flux $r^2 F_r^{(D)}$ (solid line) of magnetic helicity
per unit solid angle and the flux $[F_r^{(D)}(r) + F_{\rm tot}(r)] \, r^2$ (dashed-dotted line) of magnetic helicity
per unit solid angle which are measured in Mx$^{2}$ \, h$^{-1}$.
As follows from Fig.~\ref{Fig8}, the flux $[F_r^{(D)}(r) + F_{\rm tot}(r)] \, r^2$ (the sum of the turbulent diffusion flux
and total source flux of magnetic helicity) of small-scale field
per unit solid angle is independent of $r$, i.e.,
\begin{eqnarray}
[F_r^{(D)}(r) + F_{\rm tot}(r)] \, r^2 \approx
F_{\rm tot}(r=0.73 R_\odot) \, (0.73 R_\odot)^2 .
\label{GGM4}
\end{eqnarray}
Here we take into account that the turbulent diffusion flux $F_r^{(D)}(r=0.73 R_\odot) \to 0$ vanishes at the bottom
of the convective zone, $r=0.73 R_\odot$, where the turbulence intensity vanishes (see Fig.~\ref{Fig8}).
Equation~(\ref{GGM4}) implies that there is no accumulation of small-scale magnetic helicity
inside the solar convective zone.

\begin{figure}
   \centering
\includegraphics[width=7.5cm]{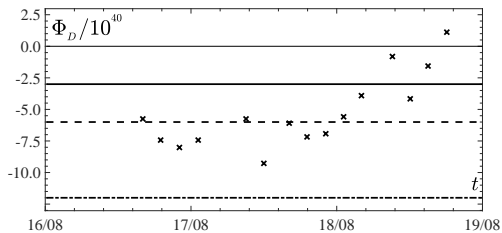}
\caption{\label{Fig9}
Comparison of the theoretical predictions for
$\Phi_D=F_r^{(D)}(r=R_\odot) \, R_\odot^2 \, \delta\phi_\ast$
with the observational values of $\Phi_D$ (slanting crosses)
which are taken from Fig.~8a by Chae et al. (2001), where
time variations of the rates of magnetic helicity change by photospheric motions (which do not include differential rotation)
are shown. Here the flux $\Phi_D$ is measured in Mx$^{2}$ \, h$^{-1}$ and $\delta\phi_\ast = 2 \pi \, \sin (\pi/4)$
is the solid angle corresponding to the thickness of the Royal sunspot region.
The theoretical values for $\Phi_D$ are given for different values of the mean magnetic
field, $\meanB_{\rm bot}$  and $\meanB_{\rm top}$, at the bottom and top of the solar convective zone
(i.e., thick solid line is for $\meanB_{\rm bot}=10^3$ G
and $\meanB_{\rm top}=8$ G; dashed line is for $\meanB_{\rm bot}=1.4 \times 10^3$ G
and $\meanB_{\rm top}=11$ G and dashed-dotted line is for $\meanB_{\rm bot}=2 \times 10^3$ G
and $\meanB_{\rm top}=16$ G).
}
\end{figure}

In Fig.~\ref{Fig9} we compare the theoretical predictions for flux
$\Phi_D \equiv F_r^{(D)}(r=R_\odot) \, R_\odot^2 \, \delta\phi_\ast$
with the observational values of $\Phi_D$
which are taken from Fig.~8a by \cite{CWQ01}, where
time variations of the rates of magnetic helicity change by photospheric motions (which do not include differential rotation)
are shown. Here the flux $\Phi_D$ is measured in Mx$^{2}$ \, h$^{-1}$ and $\delta\phi_\ast = 2 \pi \, \sin (\pi/4)$
is the solid angle corresponding to the thickness of the Royal sunspot region.
The theoretical values for $\Phi_D$ are given for different values of the mean magnetic
field, $\meanB_{\rm bot}$  and $\meanB_{\rm top}$, at the bottom and top of the solar convective zone
(see the caption of  Fig.~\ref{Fig9}).
Note that the measurements of the magnetic helicity flux are based on the equation
$\partial H_{\rm m}/ \partial t = - 2 \oint ({\bm u} \cdot {\bm a}_p) \, {\bm b}_z \, \,dS$ \citep{CWQ01,PB14},
where we use here the lower-case letters for the small-scale fields.
This implies that the measurements by \cite{CWQ01} are based on the calculation of the third-order moment,
$\langle({\bm u} \cdot {\bm a}_p) \, {\bm b}_z\rangle$,
which we describe using the turbulent diffusion approximation,
$F_r^{(D)} =  - D_{rr}^{({\rm H})} \, \nabla_r H_{\rm m}$.
As follows from Fig.~\ref{Fig9}, the theoretical predictions for flux
$\Phi_D$ are in agreement with the observational values of $\Phi_D$.

\section{Discussion and conclusions}
\label{sect5}

In the present study,  turbulent magnetic helicity fluxes of small-scale field
are derived applying the mean-field approach and the spectral $\tau$
approximation using the Coulomb gauge in a density-stratified turbulence.
The turbulent magnetic helicity fluxes contain
non-gradient contribution that is proportional to
the effective pumping velocity multiplied by the small-scale magnetic helicity.
There is the gradient contribution to the turbulent magnetic helicity flux
describing the turbulent magnetic diffusion
of the small-scale magnetic helicity.
The turbulent magnetic helicity flux includes also
the source term proportional to the kinetic $\alpha$ effect
or its radial gradient.
Finally, there is a contribution to the turbulent magnetic helicity
flux due to the solar differential rotation.

The convective zone of the Sun and solar-like stars as well as
galactic discs are the source for production of turbulent magnetic helicity fluxes.
The turbulent magnetic helicity flux due to
the kinetic $\alpha$ effect  and its radial derivative
in combination with the turbulent magnetic diffusion
of the small-scale magnetic helicity are dominant
in the solar convective zone.
The turbulent magnetic helicity fluxes result in evacuation of small-scale magnetic helicity
from the regions of generation of the solar magnetic field, which
allows to avoid the catastrophic quenching of the $\alpha$ effect.
The small-scale magnetic helicity is not accumulated inside
the solar convective zone due to turbulent magnetic diffusion flux.

The magnetic helicity fluxes are measured in the solar surface.
Most of the measurements of the magnetic helicity fluxes are performed in
active regions.
The contributions to the measured magnetic helicity flux are from both,
the solar surface and solar interiors.

\section*{Acknowledgments}

This work was partially supported by the Russian Science Foundation (grant 21-72-20067).
We acknowledge the discussions with participants
of the Nordita Scientific Program on ''Magnetic field evolution in low density
or strongly stratified plasmas", Stockholm (May -- June 2022).

\section*{Data Availability}

\noindent
There are no new data associated with this article.

\bibliographystyle{mnras}
\bibliography{Mag-hel-flux-MNRAS}

\appendix

\section{Derivation of turbulent flux of magnetic helicity}
\label{Appendix A}

In this Section we derive turbulent flux of the magnetic helicity.
We consider developed turbulence with
large fluid and magnetic Reynolds numbers, so that
the Strouhal number (the ratio of turbulent time $\tau$
to turn-over time $\ell_0/u_0$) is of the order of unity,
and the turbulent correlation time is scale-dependent, like
in Kolmogorov type turbulence.
In this case, we perform the Fourier transformation only in
${\bm k}$ space but not in $\omega$ space,
as is usually done in studies of turbulent transport in a fully developed
Kolmogorov-type turbulence.
We take into account the nonlinear terms in equations for velocity
and magnetic fluctuations and apply the $\tau$ approach.

The $\tau$ approach is a universal tool in turbulent transport
for strongly nonlinear systems that allows us to obtain closed results
and compare them with the results of laboratory experiments, observations,
and numerical simulations.
The $\tau$ approximation reproduces many
well-known phenomena found by other methods in turbulent transport
of particles and magnetic fields, in turbulent convection and stably stratified turbulent
flows for large fluid and magnetic Reynolds and P\'{e}clet numbers.

To derive equations for the turbulent fluxes of the magnetic helicity, we need expressions in a Fourier space for
the cross-helicity tensor $g_{ij}({\bm k}) = \langle u_i(t, {\bm k}) \, b_j(t, -{\bm k}) \rangle$ and the tensor $h_{ij}({\bm k}) = \langle b_i(t, {\bm k}) \, b_j(t, -{\bm k}) \rangle$ for magnetic fluctuations.
Indeed, as follows from Eq.~(\ref{PP6}), the turbulent fluxes of the magnetic helicity
depend only on the second moments $g_{ij}$ and $h_{ij}$ (except for the last two terms, $\eta \left\langle {\bm a} \times ({\bm \nabla}
\times{\bm b}) \right\rangle$ and ${\bm F}^{({\rm III})}$ which are considered separately).
Using induction equation (\ref{PP1}) for magnetic fluctuations ${\bm b}$ and the Navier-Stokes equation for velocity fluctuations ${\bm u}$ written in a Fourier space, we derive equations for the cross-helicity tensor $g_{ij}({\bm k})$ and the tensor $h_{ij}({\bm k})$ for magnetic fluctuations as
\begin{eqnarray}
&& {\partial g_{ij}({\bm k}) \over \partial t} = - \left[{\rm i} \, {\bm k} {\bm \cdot}  \overline{\bm B} - {1 \over 2} \overline{\bm B} {\bm \cdot}  {\bm \nabla}\right]
\, \Big[f_{ij}({\bm k}) - h_{ij}({\bm k})\Big]
\nonumber\\
&& \; + \hat{\cal M}^{(b)} g_{ij}^{(III)}({\bm k}) ,
\label{MM4}
\end{eqnarray}

\begin{eqnarray}
&& {\partial h_{ij}({\bm k}) \over \partial t} = {\rm i} \, \Big({\bm k} {\bm \cdot}  \overline{\bm B}\Big) \, \Big[g_{ij}({\bm k}) - g_{ji}(-{\bm k}) \Big]
\nonumber\\
&& \; + {1 \over 2} \, \Big(\overline{\bm B} {\bm \cdot}  {\bm \nabla}\Big) \, \Big[g_{ij}({\bm k}) + g_{ji}(-{\bm k}) \Big]+ \hat{\cal M}^{(b)} h_{ij}^{(III)}({\bm k}) ,
\label{MM5}
\end{eqnarray}
where in Eqs.~(\ref{MM4})--(\ref{MM5}) we neglect terms proportional to spatial derivatives of the mean magnetic field [i.e., terms $\propto {\rm O} \big(\nabla_i \overline{B}_j\big)$].
Here $f_{ij}({\bm k}) = \langle u_i(t, {\bm k}) \, u_j(t, -{\bm k}) \rangle$, and
$\hat{\cal M}^{(b)} g_{ij}^{(III)}$ and
$\hat{\cal M}^{(b)} h_{ij}^{(III)}$
are the third-order moment terms appearing due to the nonlinear terms:
\begin{eqnarray}
&& \hat{\cal M}^{(b)} g_{ij}^{(III)}({\bm k}) = - \left\langle u_i(t, {\bm k}) \, T_j^{(b)}(t, -{\bm k}) \right\rangle
\nonumber\\
&&\quad+ \left\langle {\partial u_i(t, {\bm k}) \over \partial t} \, b_j(t, -{\bm k}) \right\rangle,
\label{MM7}
\end{eqnarray}

\begin{eqnarray}
&& \hat{\cal M}^{(b)} h_{ij}^{(III)}({\bm k}) = - \left\langle b_i(t, {\bm k}) \, T_j^{(b)}(t, -{\bm k}) \right\rangle
\nonumber\\
&& \quad - \left\langle T_i^{(b)}(t, {\bm k}) \, b_j(t, -{\bm k}) \right\rangle ,
\label{MM8}
\end{eqnarray}
where
\begin{eqnarray}
T_j^{(b)}= \left[\bec{\nabla} {\bm \times} \left({\bm u} {\bm \times} {\bm b} - \left\langle {\bm u} {\bm \times} {\bm b} \right\rangle \right)\right]_j .
\label{MM9}
\end{eqnarray}

Equations~(\ref{MM4})-(\ref{MM5}) for the second moment includes the first-order spatial
differential operators applied to the third-order
moments $\hat{\cal M}^{(b)} g_{ij}^{(III)}({\bm k})$ and $\hat{\cal M}^{(b)} h_{ij}^{(III)}({\bm k})$.
A problem arises how to close the system, i.e.,
how to express the third-order moments through the lower moments, $g_{ij}$ and $h_{ij}$
denoted as $F^{(II)}$. We use the spectral $\tau$ approximation which postulates that the deviations of the third-order moments, denoted as $\hat{\cal M} F^{(III)}({\bm k})$, from the contributions to these terms afforded by a background turbulence, $\hat{\cal M} F^{(III,0)}({\bm k})$, can be expressed through the similar deviations of the second moments, $F^{(II)}({\bm k}) - F^{(II,0)}({\bm k})$ as
\begin{eqnarray}
&& \hat{\cal M} F^{(III)}({\bm k}) - \hat{\cal M} F^{(III,0)}({\bm
k}) = - {1 \over \tau_r(k)} \, \Big[F^{(II)}({\bm k})
\nonumber\\
&& \quad - F^{(II,0)}({\bm k})\Big] \,,
\label{MM10}
\end{eqnarray}
where $\tau_r(k)$ is the scale-dependent relaxation time, which can be identified with the correlation time $\tau(k)$ of the turbulent velocity field for large fluid and magnetic Reynolds numbers. The functions with the superscript $(0)$ correspond to the background turbulence with a zero mean magnetic field. Validation of the $\tau$ approximation for different
situations has been performed in various numerical
simulations
\citep{BKM04,BRR08,BRK12b,Brandenburg(2005),BS05b,BSS05,RB11,RKK11,RKB12,RKB18,HKRB12,EKLR17}.
When the mean magnetic field is zero, the turbulent electromotive force vanishes,
which implies that $g_{ij}^{(0)}({\bm k})=0$.
We also take into account magnetic fluctuations caused by a small-scale dynamo (the dynamo with a zero mean magnetic field).
Consequently, Eq.~(\ref{MM10}) reduces to $\hat{\cal M}^{(b)} g_{ij}^{(III)}({\bm k}) = - g_{ij}({\bm k}) / \tau(k)$ and $\hat{\cal M}^{(b)} h_{ij}^{(III)}({\bm k}) = - [h_{ij}({\bm k}) - h_{ij}^{(0)}({\bm k})]/ \tau(k)$.

We assume that the characteristic time of variation of the second moments $g_{ij}({\bm k})$ and $h_{ij}({\bm k})$ are substantially larger than the correlation time $\tau(k)$ for all turbulence scales. Therefore, in a steady-state Eqs.~(\ref{MM4}) and~(\ref{MM5}) yield the following formulae for the cross-helicity tensor $g_{ij}({\bm k}) = \left\langle u_i({\bm k})\, b_j(-{\bm k}) \right\rangle$, and the function $h_{ij}({\bm k})= \left\langle b_i({\bm k})\, b_j(-{\bm k}) \right\rangle$:
\begin{eqnarray}
&& g_{ij}({\bm k}) = - \tau(k) \, \biggl\{\Big[{\rm i} \, \Big({\bm k} {\bm \cdot}  \overline{\bm B}\Big) - {1 \over 2} \Big(\overline{\bm B} {\bm \cdot}  {\bm \nabla}\Big)\Big] \, \Big[f_{ij}({\bm k})
\nonumber\\
&& \quad  - h_{ij}({\bm k})\Big] - \, \meanB_j \, \Big({\rm i} \, k_n - {1 \over 2} \nabla_n \Big)\, f_{in}({\bm k})
\biggr\},
\label{MM11}
\end{eqnarray}

\begin{eqnarray}
&&  h_{ij}({\bm k})= h_{ij}^{(0)}({\bm k}) +  \tau^2(k)\, \Big({\bm k} {\bm \cdot}  \overline{\bm B}\Big) \, \biggl[2 \Big({\bm k} {\bm \cdot}  \overline{\bm B}\Big) \,  f_{ij}({\bm k})
\nonumber\\
&& \quad - k_n \, \Big(\meanB_j \, f_{in}({\bm k}) + \meanB_i \, f_{nj}({\bm k})\Big)\biggr] .
\label{MM12}
\end{eqnarray}
In Eqs.~(\ref{MM11})--(\ref{MM12}) we neglect small contributions proportional to spatial derivatives of the mean magnetic field.
Since we consider a one way coupling (i.e., we do not consider the algebraic quenching
of the turbulent fluxes of the magnetic helicity),
the correlation functions $f_{ij}$ and $h_{ij}$ in the right-hand sides of
Eqs.~(\ref{MM11})--(\ref{MM12}) should be replaced by $f_{ij}^{(0)}$ and $h_{ij}^{(0)}$, respectively.

We use the following model for the second moment, $f_{ij}^{(0)}({\bm k},{\bm R})=\left\langle u_i({\bm k}) \, u_j(-{\bm k}) \right\rangle^{(0)}$  of velocity fluctuations in density stratified and helical turbulence in a Fourier space
\citep{RKR03}:
\begin{eqnarray}
&& f_{ij}^{(0)} = {E_u(k) \over 8 \pi k^2} \, \biggl\{ \Big[(\delta_{ij} - k_{ij})
+ {{\rm i} \over k^2} \, \big(\tilde \lambda_i k_j - \tilde \lambda_j k_i\big)\Big] \left\langle {\bm u}^2 \right\rangle
\nonumber\\
&& - {1 \over k^2} \, \Big[{\rm i} \varepsilon_{ijp} \, k_p
+ (\varepsilon_{jpm} \, k_{ip} + \varepsilon_{ipm} \, k_{jp}) \tilde\lambda_m \Big]H_{\rm u} \biggr\} ,
\label{M13}
\end{eqnarray}
where $\delta_{ij}$ is the Kronecker tensor, $k_{ij}=k_i \, k_j / k^2$ and
$\tilde\lambda_m = \lambda_m - \nabla_m /2$. The energy spectrum function $E_u(k)$ of velocity fluctuations
in the inertial range of turbulence is given by $E_u(k) = (q-1)  \, k_0^{-1} \, (k / k_{0})^{-q}$,
where the exponent $q=5/3$ corresponds to the Kolmogorov spectrum,
$k_0 \leq k \leq k_{\nu}$, the wave number $k_{0} = 1 / \ell_0$, the length $\ell_0$ is the maximum scale of random motions, the wave number $k_{\nu}=\ell_{\nu}^{-1}$, the length $\ell_{\nu} = \ell_0 {\rm Re}^{-3/4}$ is the Kolmogorov (viscous) scale.
The expression for the turbulent correlation time is given by $\tau(k) = 2 \, \tau_0 \, (k / k_{0})^{1-q}$, where $\tau_0 = \ell_0 / u_{0}$ is the characteristic turbulent time. In Eq.~(\ref{M13}) we take into account inhomogeneity of the kinetic helicity.

The model for the second moment, $h_{ij}^{(0)}({\bm k},{\bm R})=\left\langle b_i({\bm k}) \, b_j(-{\bm k}) \right\rangle^{(0)}$, of  magnetic fluctuations in a Fourier space is analogous to equation~(\ref{M13})
\begin{eqnarray}
&& h_{ij}^{(0)} = {1 \over 8 \pi k^2} \, \biggl\{E_b(k)  \, (\delta_{ij} - k_{ij}) \, \left\langle {\bm b}^2 \right\rangle - {1 \over k^2} \, \biggl[{\rm i} \varepsilon_{ijp} \, k_p
\nonumber\\
&& -  {1 \over 2} \, (\varepsilon_{jpm} \, k_{ip} + \varepsilon_{ipm} \, k_{jp}) \nabla_m \biggr] \, H_{\rm c} \, \delta(k-k_0)\biggr\} ,
\label{MM13}
\end{eqnarray}
where $H_{\rm c} = \left\langle{\bm b} \,{\bm \cdot}  \,(\bec{\nabla} {\bm \times} \, {\bm b}) \right\rangle$ is the current helicity,
$E_b(k) = (q_m-1)  \, k_b^{-1} \, (k / k_{b})^{-q_m}$
is the magnetic  energy spectrum function in the range $k_b \leq k \leq k_{\eta}$, the wave number $k_{b} = 1 / \ell_b$, the length $\ell_b$ is the maximum scale of magnetic fluctuations caused by the small-scale dynamo, and the exponent $q_m=5/3$ corresponds to the Kolmogorov spectrum for the magnetic energy.
In Eq.~(\ref{MM13}) we take into account inhomogeneity of the current helicity. We also take into account that
due to the realizability condition, the current helicity of the small-scale field is located at the integral turbulence scale
\citep{KR99}.

For the integration over angles in ${\bm k}$-space we use the following integrals:
\begin{eqnarray}
&&\int_{0}^{2\pi} \, d\varphi \int_{0}^{\pi} \sin \vartheta \,d\vartheta \, k_{ij}
 = {4 \pi \over 3} \, \delta_{ij} ,
\label{C21}
\end{eqnarray}

\begin{eqnarray}
&&\int_{0}^{2\pi} \, d\varphi \int_{0}^{\pi} \sin \vartheta \,d\vartheta \, k_{ijmn}
= {4 \pi \over 15} \, \Delta_{ijmn} ,
\label{CC21}
\end{eqnarray}

\begin{eqnarray}
&&\int_{0}^{2\pi} \, d\varphi \int_{0}^{\pi} \sin \vartheta \,d\vartheta \, k_{ijmnpq}
= {4 \pi \over 105} \, \Delta_{ijmnpq} ,
\label{CCC21}
\end{eqnarray}
where
\begin{eqnarray}
&& \Delta_{ijmn} = \delta_{ij} \delta_{mn} + \delta_{im} \delta_{jn} + \delta_{in} \delta_{jm} ,
\label{KKK20}
\end{eqnarray}
\begin{eqnarray}
&& \Delta_{ijmnpq} = \Delta_{mnpq} \, \delta_{ij} +
\Delta_{jmnq} \, \delta_{ip} + \Delta_{imnq} \, \delta_{jp}
\nonumber\\
&& + \Delta_{jmnp} \, \delta_{iq} + \Delta_{imnp} \, \delta_{jq} + \Delta_{ijmn} \, \delta_{pq} - \Delta_{ijpq} \, \delta_{mn} ,
\nonumber\\
\label{KKK21}
\end{eqnarray}
and $k_{ij} = k_i \, k_j / k^2$, $k_{ijmn} = k_i \, k_j \, k_m \, k_n / k^4$ and
$k_{ijmnpq} = k_i \, k_j \, k_m \, k_n \, k_p \, k_q/ k^6$.
We also take into account that $\Delta_{ijmm} = 5 \delta_{ij}$ and $\Delta_{ijmnpp} = 7 \Delta_{ijmn}$.

For the integration over $k$ we use the following integrals for large Reynolds numbers,
Re$=u_0 \ell_0 / \nu \gg 1$:
\begin{eqnarray}
&&\int_{k_0}^{k_\nu} \tau(k) \, E_u(k) \,dk = \tau_0,
\label{MD21}
\end{eqnarray}

\begin{eqnarray}
&&\int_{k_0}^{k_\nu} {\tau(k) \, E_u(k) \over k^2} \,dk = {q-1 \over q} \, \tau_0 \, \ell_0^2,
\label{MMD21}
\end{eqnarray}

\begin{eqnarray}
&&\int_{k_0}^{k_\nu} {\tau^2(k) \, E_u(k) \over k^2} \,dk = {4(q-1) \over 3q-1} \, \tau^2_0 \, \ell_0^2,
\label{MMMD21}
\end{eqnarray}

\begin{eqnarray}
&&\int_{k_0}^{k_\nu} \tau^2(k) \, E_u(k) \,dk = {4 \over 3} \tau_0^2 .
\label{MD22}
\end{eqnarray}

Using Eqs.~(\ref{MM11})--(\ref{MD22}), and integrating in ${\bm k}$ space,
we determine various contributions  to the turbulent flux of the small-scale magnetic helicity,
see Eqs.~(\ref{RR2})--(\ref{RR10}), and Appendix~\ref{Appendix D}.
The details of the derivations of the effect of large-scale shear on turbulent fluxes of the magnetic helicity
are discussed in Appendix~\ref{Appendix C}.

\section{Derivation of equations for the second moments}
\label{Appendix B}

In this Appendix we derive Eqs.~(\ref{MM4})--(\ref{MM5}) for the cross helicity tensor $g_{ij}({\bm k}) = \langle u_i(t, {\bm k}) \, b_j(t, -{\bm k}) \rangle$ and the tensor $h_{ij}({\bm k}) = \langle b_i(t, {\bm k}) \, b_j(t, -{\bm k}) \rangle$ for magnetic fluctuations.
To this end, we perform
several calculations that are similar to the following.
We use the equation for magnetic fluctuations obtained by subtracting equation for the mean magnetic field
from the equation for the total field:
\begin{eqnarray}
{\partial {\bm b} \over \partial t} - \bec{\nabla} {\bf \times} \left({\bm u} {\bf \times} {\bm b} - \langle{\bm u} {\bf \times} {\bm b} \rangle\right) - \eta \, \Delta {\bm b} = (\overline{\bm B} {\bf \cdot} \bec{\nabla}){\bm u} -({\bm u} {\bf \cdot} \bec{\nabla})\overline{\bm B} .
\nonumber\\
\label{II3}
\end{eqnarray}
The source term, $(\overline{\bm B} {\bf \cdot} \bec{\nabla}){\bm u}$,
in the right hand side of Eq.~(\ref{II3}) in a Fourier space reads:
\begin{eqnarray}
\left[\left(\overline{\bm B} {\bf \cdot} \bec{\nabla}\right) u_j\right]_{\bm k} &=&
{\rm i} \,k_p \, \int \overline{B}_p({\bm Q}) \, u_j({\bm k}-{\bm Q}) \,d{\bm Q} ,
\label{MM41}
\end{eqnarray}
so that the induction equation for $b_j({\bm k}_2)$ in ${\bm k}$ space is given by:
\begin{eqnarray}
&& {\partial b_j({\bm k}_2) \over \partial t} = {\rm i} \,  k^{(2)}_p \, \int \overline{B}_p({\bm Q}) \,u_j({\bm k}_2-{\bm Q}) \,d{\bm Q}
\nonumber\\
&&\quad- u_n({\bm k}_2) \, {\nabla}_n \overline{B}_j + N_j^{(b)}({\bm k}_2) ,
\label{MM25}
\end{eqnarray}
where ${\bm k}^{(2)} \equiv {\bm k}_2=-{\bm k} + {\bm  K}  / 2$.
We use the identity:
\begin{eqnarray}
&& {\partial \over \partial t} \left\langle u_i({\bm k}_1,t) \,
b_j({\bm k}_2,t) \right\rangle = \left\langle{\partial u_i({\bm k}_1,t) \over \partial t} \,
b_j({\bm k}_2,t)\right\rangle
\nonumber\\
&&\quad+ \left\langle u_i({\bm k}_1,t) \, {\partial b_j({\bm k}_2,t) \over \partial t}
\right\rangle .
\label{MMMM80}
\end{eqnarray}
First we derive equation for the second term in the right hand side of Eq.~(\ref{MMMM80}).
To this end, we multiply Eq.~(\ref{MM25}) by $u_i({\bm k}_1)$ and averaging over ensemble of turbulent velocity field, where ${\bm k}_1={\bm k} + {\bm  K}  / 2$. This yields:
\begin{eqnarray}
&& \left\langle u_i({\bm k}_1) \, {\partial b_j({\bm k}_2) \over \partial t}
\right\rangle = {\rm i} \,   \left(-k_{p} + K_{p}/2\right) \, \int \,d{\bm Q} \, \overline{B}_p({\bm Q})
\nonumber\\
&& \quad \times  \left\langle u_i({\bm k}_1) \,u_j({\bm k}_2-{\bm Q}) \right\rangle
- \left\langle u_i({\bm k}_1) \, u_n({\bm k}_2) \right\rangle \, {\nabla}_n \overline{B}_j
\nonumber\\
&& \quad + \left\langle u_i({\bm k}_1) \, N_j^{(b)}({\bm k}_2) \right\rangle ,
\label{MMMM81}
\end{eqnarray}
where for brevity of notations we omit the argument $t$ in the velocity and magnetic fields.
Next, we perform in Eq.~(\ref{MMMM81}) the Fourier transformation in the large-scale variable ${\bm K}$, i.e., we use the transformation
\begin{eqnarray*}
F({\bm R}) = \int F({\bm K})\exp({\rm i} \, {\bm K} {\bm \cdot} {\bm R})\,d {\bm  K}.
\end{eqnarray*}
The first term $S_{ij}({\bm k}, {\bm R})$ in the right hand side of the obtained equation [which originates from the first term in the right hand side of Eq.~(\ref{MM25})], is given by:
\begin{eqnarray}
&& S_{ij}({\bm k}, {\bm R}) = {\rm i} \,  \int \int \overline{B}_{p}({\bm
Q})  \, \left(-k_{p} + K_{p}/2\right) \, \exp({\rm i} \, {\bm K} {\bm \cdot} {\bm R})
\nonumber\\
&& \times \langle u_i ({\bm k} + {\bm  K} / 2) u_j(-{\bm k} + {\bm  K}  / 2
- {\bm  Q}) \rangle \,d {\bm  K} \,d {\bm  Q} .
\label{MM26}
\end{eqnarray}
Next, we introduce new variables:
\begin{eqnarray}
&& \tilde {\bm k} = (\tilde {\bm k}_{1} - \tilde {\bm k}_{2}) / 2 =
{\bm k} + {\bm  Q} / 2 ,
\nonumber\\
&& \tilde {\bm K} = \tilde {\bm k}_{1}
+ \tilde {\bm k}_{2} = {\bm  K} - {\bm  Q} ,
\label{MM28}
\end{eqnarray}
where
\begin{eqnarray}
\tilde {\bm k}_{1} &=& {\bm k} + {\bm  K} / 2 , \quad
\tilde {\bm k}_{2} = - {\bm k} + {\bm  K} / 2 - {\bm  Q} .
\label{MM27}
\end{eqnarray}
Therefore, Eq.~(\ref{MM26}) in the new variables reads
\begin{eqnarray}
&&S_{ij}({\bm k}, {\bm R}) = {\rm i} \, \int \int f_{ij}\left({\bm k} + {\bm  Q} / 2, {\bm
K} - {\bm  Q}\right) \,  \overline{B}_{p}({\bm  Q})
\nonumber\\
&&  \times \left(-k_{p} + K_{p}/2\right) \, \exp{({\rm i} \, {\bm K} {\bm \cdot}
{\bm R})} \,d {\bm  K} \,d{\bm  Q} .
\label{MM29}
\end{eqnarray}
Since $ |{\bm Q}| \ll |{\bm k}| $, we use the Taylor expansion
\begin{eqnarray}
&& f_{ij}({\bm k} + {\bm Q}/2, {\bm  K} - {\bm  Q}) \simeq
f_{ij}({\bm k},{\bm  K} - {\bm  Q})
\nonumber\\
&& \quad + \frac{1}{2} {\partial f_{ij}({\bm k},{\bm  K} - {\bm Q}) \over
\partial k_s} Q_s  + O({\bm Q}^2) ,
\label{MM30}
\end{eqnarray}
and the following identity:
\begin{eqnarray}
&& \nabla_{p} [f_{ij}({\bm k},{\bm R}) \overline{B}_{p}({\bm R})] = {\rm i} \, \int
\,d {\bm  K} \, K_{p} [f_{ij}({\bm k},{\bm R}) \overline{B}_{p}({\bm R})]_{\bm  K}
\nonumber\\
&& \quad \times \exp{({\rm i} \, {\bm K} {\bm \cdot} {\bm R})}  ,
\label{MM31}
\end{eqnarray}
where
\begin{eqnarray}
&& [f_{ij}({\bm k},{\bm R}) \overline{B}_{p}({\bm R})]_{\bm  K} = \int
f_{ij}({\bm k},{\bm  K} - {\bm  Q}) \overline{B}_{p}({\bm Q}) \,d {\bm
Q} .
\nonumber\\
\label{MMM31}
\end{eqnarray}
Therefore, Eqs.~(\ref{MM29})--(\ref{MM31}) yield
\begin{eqnarray}
&& S_{ij}({\bm k},{\bm R}) \simeq \left[- {\rm i} \,({\bm k } \cdot \overline{\bm B}) +
\frac{1}{2} (\overline{\bm B} \cdot \bec{\nabla})\right] \, f_{ij}({\bm k},{\bm R})
\nonumber\\
&&\quad- \frac{1}{2} k_{p} {\partial f_{ij}({\bm k}) \over
\partial k_s} \nabla_s \overline{B}_{p} .
\label{MM32}
\end{eqnarray}
We take into account that the terms in $g_{ij}({\bm k},{\bm R})$ with
symmetric tensors with respect to the indexes "i" and "j" do not
contribute to the turbulent electromotive force because ${\cal E}_{m} =
\varepsilon_{mij} \, \int g_{ij}({\bm k},{\bm R}) \,d{\bm k}$.
In $g_{ij}({\bm k},{\bm R})$ we also neglect the second and higher derivatives over ${\bm R}$.
This procedure yields Eq.~(\ref{MM4}). Similar calculations
are performed to derive Eq.~(\ref{MM5}).

To determine various contributions to the turbulent flux of small-scale magnetic helicity, we use the following identities:
\begin{eqnarray}
&&\left(\Delta^{-1}\right)_{{\bm k}_1} = - k^{-2} \left[1 + {{\rm i} \, ({\bm k} \cdot {\bm \nabla}) \over k^{2}} \right],
\label{MMCD10}
\end{eqnarray}

\begin{eqnarray}
&&\left(\Delta^{-1}\right)_{{\bm k}_2} = - k^{-2} \left[1 - {{\rm i} \, ({\bm k} \cdot {\bm \nabla}) \over k^{2}} \right].
\label{MMD10}
\end{eqnarray}

\section{Effect of large-scale shear}
\label{Appendix C}

In this Appendix we determine the effect of large-scale shear on turbulent fluxes of the magnetic helicity.
The cross-helicity tensor $g_{ij}^{(S)}({\bm  k})  = \langle v_i ({\bm  k}) \,  b_j(-{\bm k}) \rangle$ in turbulence with large-scale shear is given by \citep{RK2004}:
\begin{eqnarray}
&& g_{ij}^{(S)}({\bm  k})  = -{\rm  i} \,\tau \, ({\bm k}
\cdot \meanBB) \,   \biggl[f_{ij}^{(S)}({\bm k}) - {h_{ij}^{(S)}({\bm  k})  \over 4 \pi \meanrho}
\nonumber\\
&&\quad + \tau \, J_{ijmn}(\meanUU)
\left(f_{mn}^{(0)}({\bm k}) - {h_{mn}^{(0)}({\bm  k})  \over 4 \pi \meanrho} \right)\biggr] ,
\label{MLS8}
\end{eqnarray}
where the effect of large-scale shear on the tensors $f_{ij}^{(S)}({\bm  k})  = \langle v_i ({\bm  k}) \,  v_j(-{\bm k}) \rangle$ and
$h_{ij}^{(S)}({\bm  k})  = \langle b_i ({\bm  k}) \,  b_j(-{\bm k}) \rangle $ is determined by
\begin{eqnarray}
&& f_{ij}^{(S)}({\bm  k})  = \tau \, I_{ijmn}(\meanUU) \, f_{mn}^{(0)}({\bm  k}) ,
\label{CS9}
\end{eqnarray}

\begin{eqnarray}
&&h_{ij}^{(S)}({\bm  k})  = \tau \, E_{ijmn}(\meanUU) \, h_{mn}^{(0)}({\bm  k})  ,
\label{S9}
\end{eqnarray}
and the tensors $I_{ijmn}(\meanUU)$, $\, E_{ijmn}(\meanUU)$
and $J_{ijmn}(\meanUU)$ are given by
\begin{eqnarray}
&& I_{ijmn}(\meanUU) = \biggl\{2 k_{iq} \delta_{mp}
\delta_{jn} + 2 k_{jq} \delta_{im} \delta_{pn} - \delta_{im}
\delta_{jq} \delta_{np}
\nonumber\\
 && \quad - \delta_{iq} \delta_{jn} \delta_{mp}  + 4 k_{pq} \delta_{im} \delta_{jn}  + \delta_{im} \delta_{jn}
k_{q} {\partial \over \partial k_{p}}
\nonumber\\
 &&\quad - {{\rm  i} \, \lambda_r \over 2 k^2} \, \biggl[\Big(k_i  \delta_{jn} \delta_{pm}
- k_j \delta_{im} \delta_{pn}\Big) \, \Big(2 k_{rq}  - \delta_{rq}\Big)
\nonumber\\
 && \quad + k_q \Big(\delta_{ip} \delta_{jn} \delta_{rm} - \delta_{im} \delta_{jp} \delta_{rn}\Big)
- 2 k_{pq}\Big(k_i  \delta_{jn} \delta_{rm}
\nonumber\\
 && \quad - k_j \delta_{im} \delta_{rn}\Big)
\biggr] \biggr\} \nabla_{p} \meanU_{q},
\label{S10}
\end{eqnarray}

\begin{eqnarray}
&& E_{ijmn}(\meanUU) = \biggl[\delta_{im} \delta_{jq} \delta_{pn}+
 \delta_{iq} \delta_{jn} \delta_{pm}
 \nonumber\\
 && \qquad+ \delta_{im} \delta_{jn}
k_{q} {\partial \over \partial k_{p}} \biggr] \, \nabla_{p} \meanU_{q} \;,
\label{S11}
\end{eqnarray}

\begin{eqnarray}
&& J_{ijmn}(\meanUU) = \biggl\{2 k_{iq} \delta_{jn}
\delta_{pm} - \delta_{iq} \delta_{jn} \delta_{pm} + \delta_{im} \delta_{jq} \delta_{pl}
\nonumber\\
 && \quad + 2 k_{pq} \delta_{im} \delta_{jn} + \delta_{im} \delta_{jn} k_{q} {\partial \over \partial k_{p}}
  - {{\rm  i} \, \lambda_r \over 2 k^2} \, \biggl[
k_i  \delta_{jn} \delta_{pm}
\nonumber\\
&& \quad \times\Big(2 k_{rq} - \delta_{rq}\Big)
+ \delta_{jn} \delta_{rm} \, \Big(k_q   \, \delta_{ip} - 2 k_i  \, k_{pq} \Big)
\biggr] \biggr\} \nabla_{p} \meanU_{q} .
\nonumber\\
\label{S12}
\end{eqnarray}
Using Eqs.~(\ref{M13})--(\ref{MD22}),  (\ref{MMCD10})--(\ref{MMD10}),
and Eqs.~(\ref{MLS8})--(\ref{S12}), and integrating in ${\bm k}$ space,
we determine various contributions  to the turbulent flux of the small-scale magnetic helicity
caused by the differential rotation, see Eq.~(\ref{RR12}) and Appendix~\ref{Appendix D}.

\section{General form of turbulent transport coefficients}
\label{Appendix D}

Applying the method described in Appendixes~\ref{Appendix A}--\ref{Appendix C},
we have determined various contributions  to the turbulent flux of the small-scale magnetic helicity.
In particular, the general form of turbulent flux of the small-scale magnetic helicity is given by
\begin{eqnarray}
&& F_i^{({\rm m})} = V_i^{({\rm H})} \, H_{\rm m} - D_{ij}^{({\rm H})} \, \nabla_j H_{\rm m}
+ N_i^{(\alpha)} \, \alpha_{_{\rm K}}
 \nonumber\\
&& \quad+ M_{ij}^{(\alpha)} \, \nabla_j \alpha_{_{\rm K}} +
F_i^{({\rm S}0)},
\label{ARR1}
\end{eqnarray}
where the turbulent transport coefficients are given below.
The turbulent pumping velocity ${\bm V}^{({\rm H})}$ of the small-scale magnetic helicity is
\begin{eqnarray}
&& {\bm V}^{({\rm H})} = - {1 \over 15} \tau_0 \, \meanV_{\rm A}^2 \, \biggl\{{\bm \lambda} +
7 {\bm \beta} ({\bm \beta} \cdot {\bm \lambda})
+ {1 \over 7} \tau_0  \,\biggl[28 \, (\meanWW \times {\bm \lambda})
\nonumber\\
&&\quad + {139 \over 2} ({\bm \beta} \cdot {\bm \lambda}) \, (\meanWW \times {\bm \beta})
- 2 {\bm Q}^{(\lambda) } + {\bm \beta} \Big( 17\, \meanWW \cdot ({\bm \beta} \times {\bm \lambda})
\nonumber\\
&& \quad+58 \,  {\bm \lambda}  \cdot {\bm Q}^{(\beta)}\Big)
 - 31 \,  {\bm Q}^{(\beta)} \, ({\bm \beta} \cdot {\bm \lambda}) - 3  \,  {\bm \lambda} ({\bm \beta}  \cdot {\bm Q}^{(\beta)})
\nonumber\\
&& \quad- 7 \,  ({\bm \beta} \times {\bm \lambda}) \, ({\bm \beta} \cdot \meanWW)\biggr]\biggr\} .
\label{RR20}
\end{eqnarray}
Here ${\bm \beta}=\meanBB/\meanB$ is the unit vector along the mean magnetic field, $\meanV_{\rm A}=\meanB/(4 \pi \meanrho\,)^{1/2}$ is the mean Alfv\'{e}n speed,
$\meanWW={\bm \nabla} \times \meanUU$ is the mean vorticity, the vectors ${\bm Q}^{(\beta)}$ and ${\bm Q}^{(\lambda)}$ are defined as $Q_i^{(\beta)}=\beta_m \, (\partial \meanU)_{mi}$ and $Q_i^{(\lambda)}=\lambda_m \, (\partial \meanU)_{mi}$, and
the gradient of the mean velocity $\nabla_i \meanU_j$ is decomposed into symmetric, $(\partial \meanU)_{ij}=(\nabla_i \meanU_j + \nabla_j \meanU_i)/2$, and antisymmetric, $\varepsilon_{ijp} \, \meanW_p/2$ parts, i.e.,
$\nabla_i \meanU_j=(\partial \meanU)_{ij} + \varepsilon_{ijp} \, \meanW_p/2$.

The total diffusion tensor $D_{ij}^{({\rm H})}$ that describes turbulent magnetic diffusion of the small-scale
magnetic helicity, reads:
\begin{eqnarray}
&& D_{ij}^{({\rm H})} = D_{T}^{({\rm H})} \, \delta_{ij}  + {1 \over 30} \tau_0 \, \meanV_{\rm A}^2 \, \biggl\{5 \delta_{ij} - 4
\beta_i \, \beta_j +  \tau_0 \,\biggl[8 \varepsilon_{ijp}
\nonumber\\
&& \times (\meanWW \cdot {\bm \beta}) \, \beta_p+ 8 \, \beta_i \,({\bm \beta} \times \meanWW)_j
  + 14 \,  \beta_j \,({\bm \beta} \times \meanWW)_i
\nonumber\\
&&+ 4 \, \varepsilon_{iqm}\, \varepsilon_{jpn} \, \beta_m\,  \beta_n \, (\partial \meanU)_{pq}
+ {1 \over 7} \, \biggl( 8(q+1) \,   (\partial \meanU)_{ij}
\nonumber\\
&&  + 2 (41+ 34 q) \, \beta_i  \, Q^{(\beta)}_j + 2 (1- 6 q) \,  \beta_j \, Q^{(\beta)}_i + (1 + 8 q)  \delta_{ij}
\nonumber\\
&&  \times   ({\bm \beta}  \cdot {\bm Q}^{(\beta)}) \biggr)\biggr]\biggr\}+ {\tau_0 \over 2}\,\biggl[\eta_{_{T}} + {8 \over 15} \tau_0 \, \meanV_{\rm A}^2 \,
\biggr] \, \varepsilon_{ijp} \, \meanW_p .
\label{RR21}
\end{eqnarray}
In derivation of Eqs.~(\ref{RR20})--(\ref{RR21}), we take into account that $H_{\rm c} = H_{\rm m}/\ell_0^2$, and we neglect small terms  $\sim {\rm O}[\ell_0^2/L_{\rm m}^2]$ with $L_{\rm m}$ being characteristic scale of spatial variations of $H_{\rm m}$.
The turbulent magnetic helicity flux also includes
the source term ${\bm N}^{(\alpha)} \, \alpha_{_{\rm K}}$
caused by the kinetic $\alpha$ effect with ${\bm N}^{(\alpha)}$ being
\begin{eqnarray}
&& {\bm N}^{(\alpha)} = - {1 \over 10} \, \ell_0^2 \, \meanB^2 \, \biggl\{{\bm \lambda}
+ {7q-2 \over q} \, \, ({\bm \beta} \cdot {\bm \lambda}) \, {\bm \beta}
+ { (q-1) \, \tau_0 \over (3q-1)}
\nonumber\\
&& \times \biggl[10 \, ({\bm \beta} \times \meanWW) \, \, ({\bm \beta} \cdot {\bm \lambda}) - 37 (\meanWW \cdot {\bm \beta}) \, ({\bm \beta} \times {\bm \lambda}) - 4 {\bm Q}^{(\lambda)}
 \nonumber\\
&& - 4 \, ({\bm \beta}  \times {\bm Q}^{(\beta, \lambda)})
+{2 \over 7} \, \biggl(19\, {\bm \beta} \, [ ({\bm \beta} \times \meanWW) \cdot {\bm \lambda}]
- 4\, {\bm Q}^{(\beta)}
\nonumber\\
&& \times  ({\bm \beta} \cdot {\bm \lambda}) - 24 \, {\bm \beta} \, ({\bm \lambda}  \cdot {\bm Q}^{(\beta)})  + 4 \, {\bm \lambda} \, ({\bm \beta}  \cdot {\bm Q}^{(\beta)}) \biggr)
\biggr] \biggr\} ,
\label{RR22}
\end{eqnarray}
where $Q_i^{(\beta, \lambda)} = ({\bm \beta} \times{\bm \lambda})_m \, (\partial \meanU)_{mi}$.
The contribution to the turbulent magnetic helicity flux,  $\propto - \, \ell_0^2 \, \meanB^2 \, {\bm \lambda} \, \alpha_{_{\rm K}}$
[see the first term in equation~(\ref{RR22})], caused by the kinetic $\alpha$ effect, has been suggested by \cite{KMR00,KMR02,KMR03a}.

The turbulent magnetic helicity flux contains also
the source term $M_{ij}^{(\alpha)} \, \nabla_j \alpha_{_{\rm K}}$
caused by the gradient $\nabla_j \alpha_{_{\rm K}}$ of the kinetic $\alpha$ effect
with $M_{ij}^{(\alpha)}$ being
\begin{eqnarray}
&& M_{ij}^{(\alpha)} = {1 \over 20 q} \, \ell_0^2 \, \meanB^2 \, \biggl\{(2q-1) \, \delta_{ij} +
(20q-23) \, \beta_i \, \beta_j
\nonumber\\
&& + {16 \, q \, (q-1) \tau_0 \over 3q-1}
\, \biggl[\beta_i \,({\bm \beta} \times \meanWW)_j
+ (\meanWW \cdot {\bm \beta})\, \varepsilon_{ijp} \, \beta_p\biggr]\biggr\} .
\nonumber\\
\label{RR23}
\end{eqnarray}
The  additional contribution ${\bm F}^{({\rm S}0)}$ to the turbulent magnetic helicity flux
caused by the large-scale shear (differential rotation) is given by
\begin{eqnarray}
&& {\bm F}^{({\rm S}0)} =- {q-1 \over 3(q+1)} \, \ell^2_b \, \left\langle {\bm b}^2 \right\rangle \,  \meanWW + {2 \over 45} \, \ell_0^2 \, \meanB^2 \, \Big[ 11 \epsilon \,  \meanWW
\nonumber\\
&& + (3 \epsilon-10) \, ({\bm \beta} \cdot \meanWW) \, {\bm \beta} +  ({\bm \beta} \times{\bm Q}^{(\beta)})
[8q +35
\nonumber\\
&& + \epsilon(8q -20) ] \,\Big] .
\label{PPFF25}
\end{eqnarray}
Here $\epsilon = \ell^2_b \, \left\langle {\bm b}^2 \right\rangle /  (\ell^2_0 \, 4 \pi \meanrho \, \left\langle {\bm u}^2 \right\rangle)$,
and $\ell_b$ is the energy containing scale of magnetic fluctuations with a zero mean magnetic field.
The contribution to the turbulent magnetic helicity flux,  $\propto \, \ell_0^2 \, \meanB^2 \, ({\bm \beta} \times{\bm Q}^{(\beta)})$
[see the last term in equation~(\ref{PPFF25})], caused by the large-scale shear, has been derived  by
\cite{BSS05}, using a general expression originally suggested by \cite{VC01}.

To derive equations for the turbulent magnetic helicity flux due to the differential rotation
in spherical coordinates, we use the identities given below.
The large-scale shear velocity $\meanUU = {\bm \delta\Omega} \times {\bm r}$ is
caused by  the differential (non-uniform) rotation, that is in spherical coordinates $(r, \vartheta, \varphi)$ reads
\begin{eqnarray}
{\bm \delta\Omega}=\delta\Omega(r,\vartheta) \, (\cos \vartheta, - \sin \vartheta, 0) ,
\label{S9}
\end{eqnarray}
and the stress tensor $(\partial \meanU)_{ij}$ reads
 \begin{eqnarray}
(\partial \meanU)_{ij} = {r_n \over 2} \, \left(\varepsilon_{imn} \nabla_j + \varepsilon_{jmn} \nabla_i\right) \delta\Omega_m .
\label{S10}
\end{eqnarray}
The vectors ${\bm Q}^{(\beta)}$ and ${\bm Q}^{(\lambda)}$ defined as $Q_i^{(\beta)}=\beta_m \, (\partial \meanU)_{mi}$ and $Q_i^{(\lambda)}=\lambda_m \, (\partial \meanU)_{mi}$, are given by
\begin{eqnarray}
{\bm Q}^{(\beta)} &=& ({\bm r} \times {\bm \beta})_m \, ({\bm \nabla} \delta\Omega_m) - {\bm r} \times  ({\bm \beta} \cdot {\bm \nabla})
{\bm \delta\Omega} ,
\label{S8}
\end{eqnarray}

\begin{eqnarray}
{\bm Q}^{(\lambda)} &=& - {\bm r} \times  ({\bm \lambda} \cdot {\bm \nabla}) {\bm \delta\Omega} ,
\label{S7}
\end{eqnarray}
where ${\bm \lambda} = \lambda \, {\bm e}_r$ and ${\bm \beta}=\meanBB/\meanB=(\beta_r, \beta_\vartheta, \beta_\varphi)$.
We also use the identity
\begin{eqnarray}
&&  \varepsilon_{iqm}\, \varepsilon_{jpn} \, \beta_m\,  \beta_n \, (\partial \meanU)_{pq} =
 {1 \over 2} \,  ({\bm r} \cdot {\bm \beta}) \, \Big[ ({\bm \beta} \times  {\bm \nabla})_i \, \delta\Omega_j
\nonumber\\
&& \quad +  ({\bm \beta} \times  {\bm \nabla})_j \, \delta\Omega_i \Big]
- {1 \over 2} \, \beta_m \, \Big[ r_i \, ({\bm \beta} \times  {\bm \nabla})_j
\nonumber\\
&& \quad
+  r_j\, ({\bm \beta} \times  {\bm \nabla})_i  \Big] \, \delta\Omega_m .
 \label{S6}
\end{eqnarray}
We have taken into account that $ \left({\bm \beta} \times {\bm Q}^{(\beta)}\right)_r = {\rm O} ({\bm \nabla} \delta\Omega)$, i.e it
does not contain contributions $\propto \delta\Omega$, but it includes their spatial derivatives, ${\bm \nabla} \delta\Omega$.
Using Eqs.~(\ref{ARR1})--(\ref{S6}), we determine various contributions  to the turbulent flux of the small-scale magnetic helicity
in spherical coordinates, see Eqs.~(\ref{SRR1})--(\ref{RR12}).

\section{Turbulent transport coefficients in the Cartesian coordinates}
\label{Appendix E}

For better understanding of the physics related to various contributions
to the turbulent flux of the small-scale magnetic helicity
[see Eqs.~(\ref{ARR1})--(\ref{S6})], we consider a small-scale turbulence with large-scale
linear velocity shear $\meanUU=(0, Sx, 0)$ in the Cartesian coordinates.
In this case, the large-scale vorticity is $\meanWW=(0, 0, S)$, the stress tensor
$(\partial \meanU)_{ij} = (S/2) \, (e_i^x \, e_j^y + e_j^x \, e_i^y)$, the vector ${\bm \lambda}$ that describes the non-uniform mean fluid density, is ${\bm \lambda}=\lambda \, (\sin \vartheta, 0, \cos \vartheta)$, the unit vector along the large-scale magnetic  is ${\bm \beta}=(\cos \tilde \beta, \sin \tilde \beta, 0)$, the vector $Q_i^{(\beta)}=\beta_m \, (\partial \meanU)_{mi}= (S/2) \, (\sin \tilde \beta, \cos \tilde \beta, 0)$ and the vector $Q_i^{(\lambda)}=\lambda_m \, (\partial \meanU)_{mi}= (\lambda \, S/2) \, \sin \vartheta \,  e_i^y$.
We also take into account that
\begin{eqnarray}
&& {\bm \beta} \times {\bm \lambda}= \lambda \, (\cos \vartheta \, \sin \tilde \beta, - \cos \vartheta \, \cos \tilde \beta, - \sin \vartheta \, \sin \tilde \beta),
\label{PMM10}
\end{eqnarray}

\begin{eqnarray}
&& ({\bm \beta}  \times {\bm Q}^{(\beta)})_i = (S/2) \, \cos (2 \tilde \beta) \,  e_i^z,
\label{PMM11}
\end{eqnarray}

\begin{eqnarray}
&& ({\bm \beta}  \times {\bm Q}^{(\lambda)})_i = (S \, \lambda/2) \, \sin \vartheta \, \cos \tilde \beta \,  e_i^z,
\label{PMM12}
\end{eqnarray}

\begin{eqnarray}
&& {\bm \beta} \times {\bm W}= S \, (\sin \tilde \beta, - \cos \tilde \beta, 0),
\label{PMM13}
\end{eqnarray}

\begin{eqnarray}
&& (\meanWW \times {\bm \lambda})_i = S \, \lambda \, \sin \vartheta \, e_i^y .
\label{PMM14}
\end{eqnarray}

First, we determine various contributions to the turbulent flux of the magnetic helicity inside the turbulent region
where the toroidal mean magnetic field is much larger than the poloidal mean magnetic field, i.e., ${\bm \beta}=(0, 1, 0)$.
In this case, the turbulent pumping velocity ${\bm V}^{({\rm H})}$ of the small-scale magnetic helicity is
\begin{eqnarray}
{\bm V}^{({\rm H})} &=& - {1 \over 15} \tau_0 \, \meanV_{\rm A}^2 \,  \lambda \, \biggl[\biggl(1 + {3 \over 14} S  \,\tau_0
\biggr) \, {\bm e}_ \lambda + 5.6 \, S \, \tau_0  \,   {\bm e}^y \biggr],
\nonumber\\
\label{LL20}
\end{eqnarray}
where ${\bm e}_ \lambda={\bm \lambda}/\lambda$.
The turbulent magnetic helicity flux has
the source term ${\bm N}^{(\alpha)} \, \alpha_{_{\rm K}}$
caused by the kinetic $\alpha$ effect with ${\bm N}^{(\alpha)}$ being
\begin{eqnarray}
&& {\bm N}^{(\alpha)} = - {1 \over 10} \, \ell_0^2 \, \meanB^2 \, {\bm \lambda} \, \biggl[1 - { 4(q-1) \over 7 (3q-1)} \, S  \,\tau_0
\biggr] .
\label{LL21}
\end{eqnarray}
The total diffusion tensor $D_{ij}^{({\rm H})}$
which describes the microscopic and turbulent magnetic diffusion of the small-scale
magnetic helicity is given by:
\begin{eqnarray}
&& D_{ij}^{({\rm H})} = D_1 \, \delta_{ij} - D_2 e_i^y \, e_j^y + D_3 e_i^x \, e_j^y
- D_4 e_i^y \, e_j^x ,
\label{LL23}
\end{eqnarray}
where $D_2 = (2/15) \, \tau_0 \, \meanV_{\rm A}^2$,
\begin{eqnarray}
D_1 &=& D_{T}^{({\rm H})} + {1 \over 3} \eta + {1 \over 6} \tau_0 \, \meanV_{\rm A}^2
 \, \biggl[1 - {1 + 8q \over 70} \, S  \,\tau_0 \biggr] ,
\label{LL24}
\end{eqnarray}

\begin{eqnarray}
D_3 &=&{1 \over 2} \, S  \,\tau_0  \, \biggl[\eta_{_{T}} + {159-6q \over 105} \, \tau_0 \, \meanV_{\rm A}^2 \biggr] ,
\label{LL25}
\end{eqnarray}

\begin{eqnarray}
D_4 &=& {1 \over 2} \, S  \,\tau_0  \, \biggl[\eta_{_{T}} - {34q + 45\over 105} \, \tau_0 \, \meanV_{\rm A}^2 \biggr] .
\label{LL26}
\end{eqnarray}
Equation~(\ref{LL23}) implies that $D_{xx}^{({\rm H})}=D_{zz}^{({\rm H})}=D_1$,
$D_{yy}^{({\rm H})} = D_1-D_2$, $D_{xy}^{({\rm H})} = D_3$, $D_{yx}^{({\rm H})} = - D_4$,
and other components of the total diffusion tensor $D_{ij}^{({\rm H})}$ vanish.
The turbulent magnetic helicity flux containing
the source term $M_{ij}^{(\alpha)} \, \nabla_j \alpha_{_{\rm K}}$
with $M_{ij}^{(\alpha)}$ being
\begin{eqnarray}
&& M_{ij}^{(\alpha)} = {1 \over 20 q} \, \ell_0^2 \, \meanB^2 \, \biggl[(2q-1) \, \delta_{ij} +
(20q-23) \, e_i^y \, e_j^y
\nonumber\\
&& \quad + {16 \, q \, (q-1) \over 3q-1} \, S \, \tau_0 \, e_i^y \, e_j^x\biggr] .
\label{LL27}
\end{eqnarray}
The  additional contribution ${\bm F}^{({\rm S}0)}$ to the turbulent magnetic helicity flux
caused by the large-scale shear is given by
\begin{eqnarray}
&& {\bm F}^{({\rm S}0)} =- \biggl[{q-1 \over 3(q+1)}  - {22 \over 45} \, {\meanV_{\rm A}^2
\over \left\langle {\bm u}^2 \right\rangle} \,  \biggr] \, \ell^2_b \, \left\langle {\bm b}^2 \right\rangle \, S \, {\bm e}^z.
\label{LL22}
\end{eqnarray}

Now we determine various contributions to the turbulent flux of the magnetic helicity
at the surface (the upper boundary of the turbulent region),
where the toroidal mean magnetic field is much smaller than the poloidal mean magnetic field, i.e., ${\bm \beta}=(1, 0, 0)$.
In this case, the turbulent pumping velocity ${\bm V}^{({\rm H})}$ of the small-scale magnetic helicity is
\begin{eqnarray}
{\bm V}^{({\rm H})} &=& - {1 \over 15} \tau_0 \, \meanV_{\rm A}^2 \, \lambda \, \biggl[{\bm e}_ \lambda + 7 \, \sin \vartheta \, \biggl({\bm e}^x + {81 \over 49} \, S  \,\tau_0 \, {\bm e}^y\biggl) \biggr] .
\nonumber\\
\label{LL30}
\end{eqnarray}
The turbulent magnetic helicity flux has
the source term ${\bm N}^{(\alpha)} \, \alpha_{_{\rm K}}$
caused by the kinetic $\alpha$ effect with ${\bm N}^{(\alpha)}$ being
\begin{eqnarray}
&& {\bm N}^{(\alpha)} = - {1 \over 10} \, \ell_0^2 \, \meanB^2 \, \lambda\, \biggl[{\bm e}_ \lambda + {7q-2\over q}
\, \sin \vartheta \, {\bm e}^x
\nonumber\\
&& \quad - {2(q-1) \over 3q-1} \,  S  \,\tau_0 \, \biggl({\bm e}^z + {44 \over 7}  \, \sin \vartheta \,{\bm e}^y\biggl) \biggr] .
\label{LL31}
\end{eqnarray}
The total diffusion tensor $D_{ij}^{({\rm H})}$
which describes the microscopic and turbulent magnetic diffusion of the small-scale
magnetic helicity is given by:
\begin{eqnarray}
&& D_{ij}^{({\rm H})} = D_1 \, \delta_{ij} - D_2 e_i^x \, e_j^x + D_3 e_i^x \, e_j^y
- D_4 e_i^y \, e_j^x ,
\label{LL33}
\end{eqnarray}
where $D_2 = (2/15) \, \tau_0 \, \meanV_{\rm A}^2$,
\begin{eqnarray}
D_1 &=& D_{T}^{({\rm H})} + {1 \over 3} \eta + {1 \over 6} \tau_0 \, \meanV_{\rm A}^2,
\label{LL34}
\end{eqnarray}

\begin{eqnarray}
D_3 &=&{1 \over 2} \, S  \,\tau_0  \, \biggl[\eta_{_{T}} + {49+ 42q \over 105} \, \tau_0 \, \meanV_{\rm A}^2 \biggr] ,
\label{LL35}
\end{eqnarray}

\begin{eqnarray}
D_4 &=& {1 \over 2} \, S  \,\tau_0  \, \biggl[\eta_{_{T}} + {145 - 2q\over 105} \, \tau_0 \, \meanV_{\rm A}^2 \biggr] .
\label{LL36}
\end{eqnarray}
Equation~(\ref{LL33}) implies that $D_{yy}^{({\rm H})}=D_{zz}^{({\rm H})}=D_1$,
$D_{xx}^{({\rm H})} = D_1-D_2$, $D_{xy}^{({\rm H})} = D_3$, $D_{yx}^{({\rm H})} = - D_4$,
and other components of the total diffusion tensor $D_{ij}^{({\rm H})}$ vanish.
The turbulent magnetic helicity flux containing
the source term $M_{ij}^{(\alpha)} \, \nabla_j \alpha_{_{\rm K}}$
with $M_{ij}^{(\alpha)}$ being
\begin{eqnarray}
&& M_{ij}^{(\alpha)} = {1 \over 20 q} \, \ell_0^2 \, \meanB^2 \, \biggl[(2q-1) \, \delta_{ij} +
(20q-23) \, e_i^x \, e_j^x
\nonumber\\
&& \quad - {16 \, q \, (q-1) \over 3q-1} \, S \, \tau_0 \, e_i^x \, e_j^y\biggr] .
\label{LL37}
\end{eqnarray}
The additional contribution ${\bm F}^{({\rm S}0)}$ to the turbulent magnetic helicity flux
caused by the large-scale shear is given by
\begin{eqnarray}
&& {\bm F}^{({\rm S}0)} =  {1 \over 3} \biggl[{8q +35 \over 15} \, \ell_0^2 \, \meanB^2
- \ell^2_b \, \left\langle {\bm b}^2 \right \rangle  \, \biggl({q-1 \over q+1}
\nonumber\\
&& \quad - {2 \, (4q +1)  \over 15} \, {\meanV_{\rm A}^2  \over \left\langle {\bm u}^2 \right\rangle}  \biggr) \biggr] \, S \,{\bm e}^z .
\label{LL32}
\end{eqnarray}

\bsp	
\label{lastpage}
\end{document}